\documentclass{article}
\usepackage{spconf,amsmath,graphicx}

\title{A NEW METHOD FOR ESTIMATING THE WIDTHS OF JPEG IMAGES}
\name{Xianyan Wu, Qi Han$^*$, Dan Le, Xiamu Niu\thanks{We would like to give many thanks to Qiang Zhao and Jinghua Yu for their helpful suggestions about writing. This work is supported by the National Natural Science Foundation of China(Grant Number:61361166006, 61100187).}}
\address{School  of Computer Science and Technology, Harbin Institute of Technology, Harbin 150001, China\\
wuxianyan11@gmail.com, qi.han@hit.edu.cn, ledan@hit.edu.cn, xiamu.niu@hit.edu.cn
}
\begin{document}
%
\maketitle
\begin{abstract}
Image width is important for image understanding. We propose a novel method to estimate widths for JPEG images when their widths  are not available. The key idea is that the distance between two decoded MCUs (Minimum Coded Unit) adjacent in the vertical direction is usually small, which is measured by the average Euclidean distance between the pixels from the bottom row of the top MCU and the top row of the bottom MCU. On PASCAL VOC 2010 challenge dataset and USC-SIPI image database, experimental results
show the high performance of the proposed approach.


%
%

\end{abstract}
\section{Introduction}
\label{sec:intro}



File carving \cite{pal2009evolution, memon2006automated, cohen2008advanced, poisel2011advanced, xu2009reassembling} is a forensic technique that recovers files based merely on file structure and content and without any matching file system meta-data. A good file carving technique is important for the forensic investigators, because they need its help to seize digital evidence to combat computer-related crimes, such as illegal hacking of computers, identity theft, child pornography and child grooming. The digital evidence can be stored in kinds of digital files, such as  network log, text document, video, and image, where image is one of the most common. Since JPEG is the most widely adopted image format, which also is the default format used by the majority of digital cameras and smartphones, JPEG file carving \cite{cohen2008advanced,karresand2008reassembly,xu2009reassembling,huang2013method} has aroused numerous researchers' great interest.


The earliest file carving techniques use "magic numbers"  which are the byte sequences identifying the file header and footer. For example, the magic numbers of JPEG images are hex sequences FFD8 and FFD9.  They firstly identify the file header and footer by the "magic numbers". The file header identifies the starting bytes of a file while the file footer identifies the ending bytes of the file. Secondly,  these methods merge and return all unallocated clusters between the file header and footer. It is noted that the "magic numbers" identifying the file footer can be replaced by other information like the file size for some file types, such as the Windows BMP file, in which case they will merge and return as many unallocated sequential clusters following the header as required to equal the file size. Foremost\footnote{http://foremost.sourceforge.net} is one of the earliest carving techniques employed "magic numbers", whose performance and memory usage were improved by Golden et al \cite{richard2005scalpel}.

For all the earliest file carving techniques, the underlying assumptions are that the "magic numbers" must be available and the file data must be stored in the correct sequence on consecutive clusters on disk. Unfortunately, the underlying assumptions are not always true. On one hand, the "magic numbers" do not be contained in many file formats at all, or may be unavailable because the part of file containing them are missing. On the other hand, some files are fragmented and not stored in the correct sequence on consecutive clusters on disk. According to Garfinkel's fragmentation statistics \cite{garfinkel2007carving} which is collected from over 350 disks containing FAT, NTFA, and Unix file systems, the fragmentation rate of file types likely to be of interest by forensic examiners are high, such as 22\%, 17\%, 16\% and 58\% for  AVI, DOC, JPEG and PST respectively, although fragmentation in a typical disk is low.

%
%
%

To overcome the above problems, many algorithms to recover the fragmented files are proposed. The recovery process mainly consists of three phases: preprocessing, collation and reassembly. Preprocessing refers as preliminary work, such as decrypt encrypted devices. Collation is file fragmented type classification, for which the typical methods are seen \cite{beebe2013sceadan,mcdaniel2003content,li2005fileprints,gopal2011statistical,axelsson2010normalised,penrose2013approaches}. Reassembly  is  reordering identified fragments. One of the first works to reassemble the fragmented files is proposed by Shanmugasundaram et al. \cite{shanmugasundaram2002automatic}. They cast  the problem of recovery as a Hamiltonian path problem solved by the alpha-beta heuristic from game theory. The approaches for reassembling the fragmented JPEG files can be mainly divided into the following groups: based on a mapping function\cite{cohen2007advanced,cohen2008advanced}, SmartCarver \cite{pal2009evolution}, graph-based\cite{shanmugasundaram2002automatic,pal2003automated,memon2006automated}, Bifragment gap carving\cite{garfinkel2007carving}, based on restart markers \cite{karresand2008reassembly} and based on thumbnails\cite{mohamad2010mykarve,guo2011method}. The problem of those methods is that they work well only for the files with an available header, because the header can afford some important information for decoding, such as the number of the color components, sampling factors for all components, Huffman tables, quantization tables and image width. If any of the first three is wrong, the decoding will fail.
Although the last two will not affect successful decoding, they will make the decoded image visually degraded and image blocks shifted respectively. Therefore it's necessary to estimate all these information when they are missing.

%


Huffman tables obtain the researchers' attention firstly. Husrev et al \cite{sencar2009identification} proposes to obtain Huffman tables by sharing with the recovered files stored on the same medium, because these image files may be captured by the same camera, edited by the same software tools, or downloaded from the same Web pages, making a relation may exist between them. When combined with the approach described in Section 3 in \cite{sencar2009identification}, it possible to identify whether or not the stand-alone fragment is generated by a set of Huffman tables of other recovered images or by any of the known Huffman tables. Karresand et al \cite{karresand2008reassembly} have checked images from 76 popular digital cameras and found that 69(91\%) of them use the same stand Huffman tables, four cameras own unique Huffman tables, and the other three have the same but non-standard tables.  Hence they point out a brute force model is feasible. For the number of the color components and sampling factors for all components,they both can be gained by enumerating, because there are only 3 possible values for color components number and 12 different possible combinations of sampling factors \cite{karresand2008reassembly}.
 For the quantization tables, \cite{huang2013method} have proposed an alternative method to approximately reconstruct them via the saturated overflow.

\begin{figure}[t]
\begin{minipage}[b]{0.45\linewidth}
  \centering
  \centerline{\includegraphics[width=4.3cm]{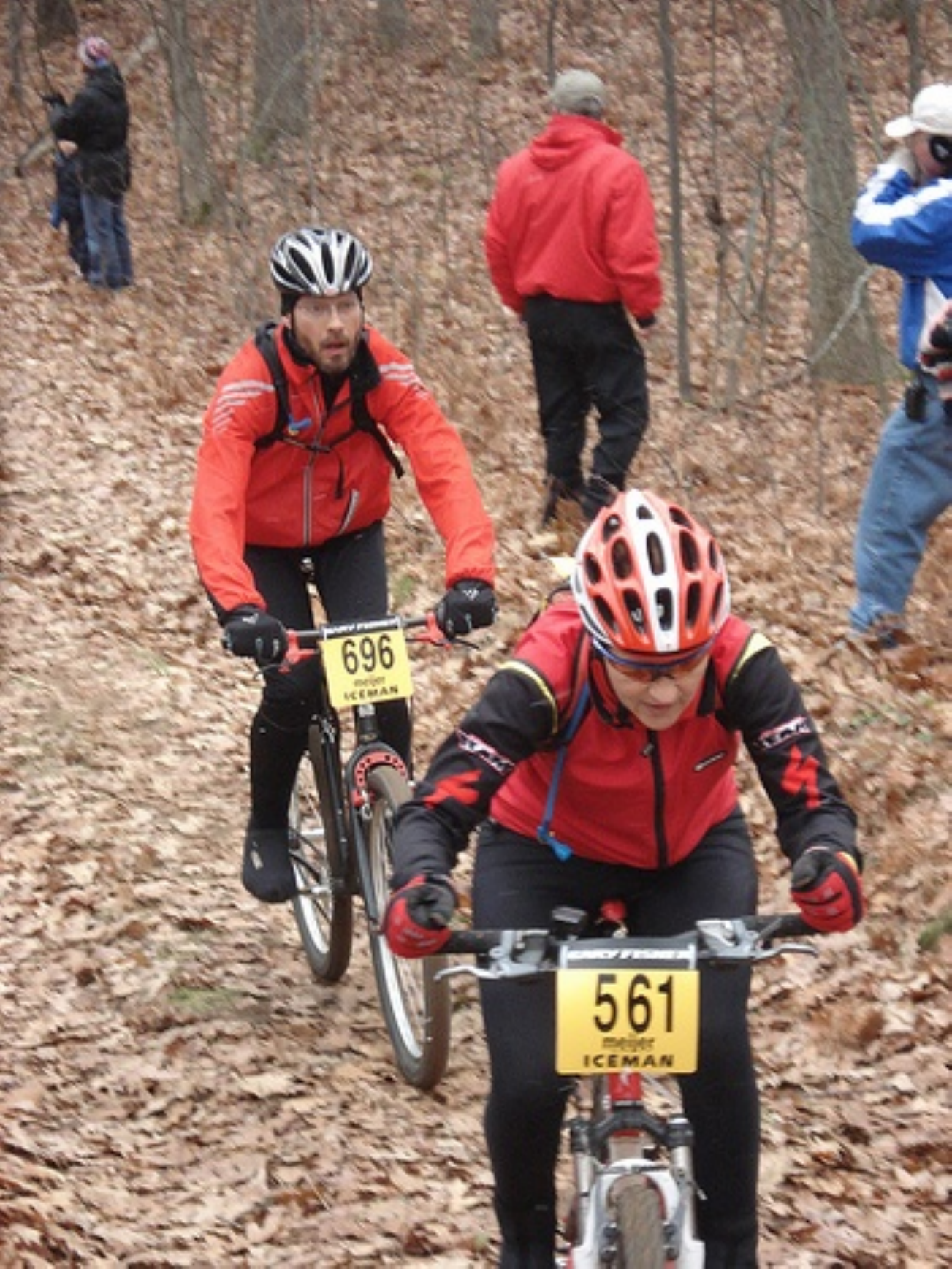}}
  \vspace{0.07cm}
  \centerline{(a)}\medskip
\end{minipage}
\hfill
\begin{minipage}[b]{0.45\linewidth}
  \centering
  \centerline{\includegraphics[width=4.3cm]{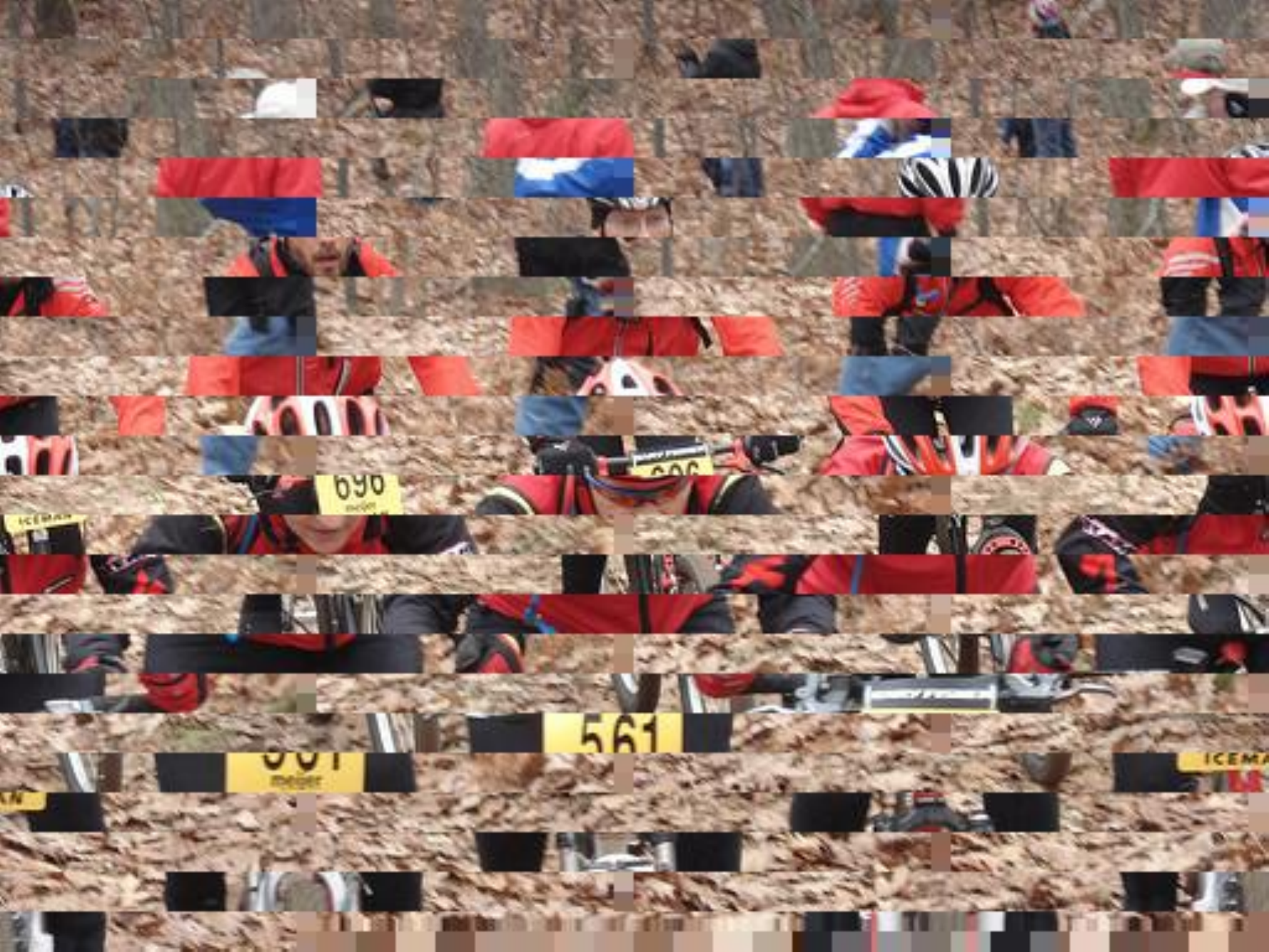}}
  \vspace{0.07cm}
  \centerline{(b)}\medskip
 \end{minipage}
%

\caption{An image from PASCAL VOC2010 is decoded with wrong width. (a) The original image \emph{2007\_001630.jpg} with width 375. (b) The image is decoded with the wrong width 512.}

\label{fig:The images showed by wrong width}

\end{figure}

Based on the rate of change of DC values, \cite{karresand2008reassembly} proposes a method to estimate the image width to help to reassemble fragmented baseline sequential JPEG images containing restart markers. However the method works under the assumption that there is at least one vertically oriented line in the test image. In their experiments, the widths estimation for some images are not correct, and the incorrect widths just affect reassembly of a few images.
 But the incorrect widths will have a strong impact on image understanding, see Fig.\ref{fig:The images showed by wrong width}.
 But the incorrect widths will have a strong impact on image understanding, see Fig.\ref{fig:The images showed by wrong width}.
 To improve this method, \cite{xu2010width} uses a differential algorithm to measure the similarities between rows of DC luminance values in specified $S$, where $S$ is the number of DC coefficients per row. Then the image width is the product of 8 and ${S^*}$ which minimizes the similarities. The performance of the algorithm depends on the interval of $S$. However the presented method to set the interval is not clear. Besides, just an image is tested, which is randomly downloaded from internet with the size of 800*600, and no results for any other images. So the effectiveness of their method can not be sufficiently verified.

%
%

 We will present a novel approach to estimate the image width. The key intuition is that the distance between two decoded MCUs adjacent in the vertical direction is small, which is measured by the average Euclidean distance between the pixels from the bottom row of the top MCU and the top row of the bottom MCU. In order to fully evaluate the proposed method, we employ two popular image databases to verify the performance of our method.  Experimental results show that the widths of almost all the test images can be correctly estimated.


The rest of this paper is organized as follows: we present our method in section \ref{sec:PROPOSED METHOD}. Section \ref{sec:EXPERIMENTS} gives an experimental evaluation, followed by conclusions in section \ref{sec:CONCLUSION}.

\section{PROPOSED METHOD}
\label{sec:PROPOSED METHOD}


We assume that the JPEG data can be decoded successfully into MCU sequences with correct order.  When the information important for decoding are missing, they can be obtained by some ways as stated in section \ref{sec:intro},  making the existing approaches for reassembling the fragmented JPEG files can works well even if these important information have been missing. Hence, we can always gain MCU sequences with correct order. And our assumption is reasonable.

\begin{figure}[htbp]
\includegraphics[width=0.45\textwidth]{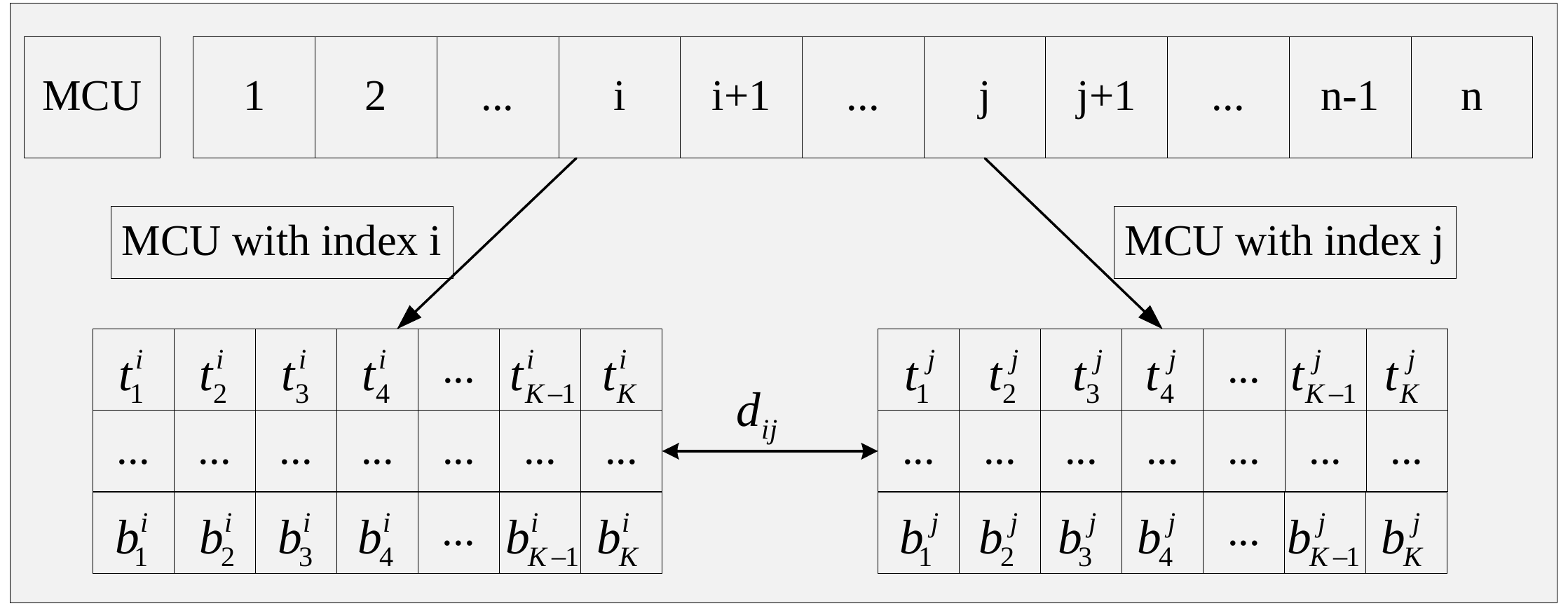}
\caption{\label{MCU} Top row: the diagram of all $n$ MCUs decoded from test image $I$ and their correct order is $1,2,...,n$. Bottom left: the diagram of the MCU with index $i$. Bottom right: the diagram of the MCU with index $j$.}
\end{figure}

Under our assumption, the JPEG data have been decoded successfully into MCU sequences with correct order which are numbered by $1,2,...,n$, as shown in the top row of Fig.\ref{MCU}. Since a JPEG file in sequential mode\footnote{Sequential mode is the most common JPEG mode,and we only focus on this mode in this paper.}  is stored as one top-to-bottom scan of the image, the image width is equal to
\begin{equation}
(j-i)*K
\label{widthbyK}
\end{equation}
where $i$ and $j$ are the indexes of the two MCUs adjacent in the vertical direction $(i<j)$, and $K$ is MCU width, i.e. the number of the pixels in the horizontal direction of MCUs, which remains the same for any given JPEG file. Hence the key point to obtain image width is to find the pairs of  MCUs adjacent in the vertical direction most likely.

According to the local similarity of images, two MCUs adjacent in the vertical direction are similar. We employ the average Euclidean
distance between the pixels from the bottom row of the top MCU and the top row of the bottom MCU to measure the local similarity. For brevity, see Fig.\ref{MCU}, let the MCU with index $i$ be the top one, the MCU with index $j$ be the bottom one, the top row  pixels of the MCU with index $i$ be  ${t_i} = (t_1^i,t_2^i, \cdots ,t_K^i)'\;\;$, and the bottom row  pixels of the MCU with index $i$ be ${b_i} = (b_1^i,b_2^i, \cdots ,b_K^i)'\;$, where $t_k^i = (t_{k1}^i,t_{k2}^i,...,t_{kC}^i)$ and $b_k^i = (b_{k1}^i,b_{k2}^i,...,b_{kC}^i)$ both represent a pixel, and $C$ is the number of the color components.
Then the distance between the MCU with index $i$ and the MCU with index $j$ is
\begin{equation}
{d_{ij}} = \frac{1}{K}\sum\limits_{k = 1}^{k = K} {\sqrt {\sum\limits_{c = 1}^C {{{(b_{kc}^i - t_{kc}^j)}^2}} } }, \; i < j.
\label{distance}
\end{equation}
The smaller the distance is, the more possible they are adjacent in vertical direction.

For each MCU with index $i(1\le i<n-1)$, we find out the MCUs most likely adjacent to it in the vertical direction from the MCUs with index
$\left\{ {i + 1,i + 2, \cdots ,n} \right\}$.  According to Eq.(\ref{widthbyK}), these candidate MCUs pairs are used to calculate candidate widths. The one which appears with the highest frequency  among  all the candidate widths is chosen as the image width. Specifically, the proposed method includes the following steps:

Step $1$, for each MCU with index $i(1\le i<n)$, calculate the distance ${d_{ij}}$  between it and the MCUs with index $j$($ i<j\le n$) according to Eq.(\ref{distance}). These distances are denoted as the following array
\begin{equation}
{D_i} = \left[ {{d_{i(i + 1)}},{d_{i(i + 2)}}, \cdots ,{d_{in}}} \right].
\label{distance_all}
\end{equation}

Step $2$, for each MCU with index $i(1\le i<n)$, find out the candidate MCUs most likely adjacent to it in the vertical direction, i.e. find out the indexes of the minimum of ${D_i}$ as follows,
\begin{equation}
{M_i} = \left\{ {\left. j \right|{d_{ij}} = \min \left\{ {{D_i}} \right\},i < j \le n} \right\}.
\label{minimum_points index a}
\end{equation}


Step $3$, for each MCU with index $i(1\le i<n)$, according to Eq.(\ref{widthbyK}) and  ${M_i}$, calculate the candidate widths as follows,
\begin{equation}
{W_i} = \left\{ {\left. {{w_{ij}}} \right|{w_{ij}} = \left( {j - i} \right) \cdot K,\;\forall j \in {M_i}} \right\},
\label{compute pixel interval}
\end{equation}
 where $K \in S$ is MCU width and $ S = \{ 8,16\}$ is the set containing all the possible values for $K$. The specific value for $K$ is determined by the sampling factor.


Step $4$, splice all the candidate widths in ${W_i}(1\le i <n)$ into a bigger array $W$ as follows,
\[W = \left[ {{W_1},{W_2}, \cdots ,{W_{n - 1}}} \right].\]
Besides, count the frequency of each candidate width in $W$ as follows,
\[\left( {F,V} \right) = {\mathop{\rm COUNT}\nolimits} \left( W \right),\]
where $F$ is the frequencies array and $V$ is the candidate widths array.

Step $5$,  choose the candidate width with the highest frequency as the image width
\[\begin{array}{l}
\hat i = \mathop {\arg \max }\limits_i \left\{ F \right\}\\
{w^ * } = V\left( {\hat i} \right)
\end{array}.\]

\begin{figure}[htbp]
\includegraphics[width=0.5\textwidth]{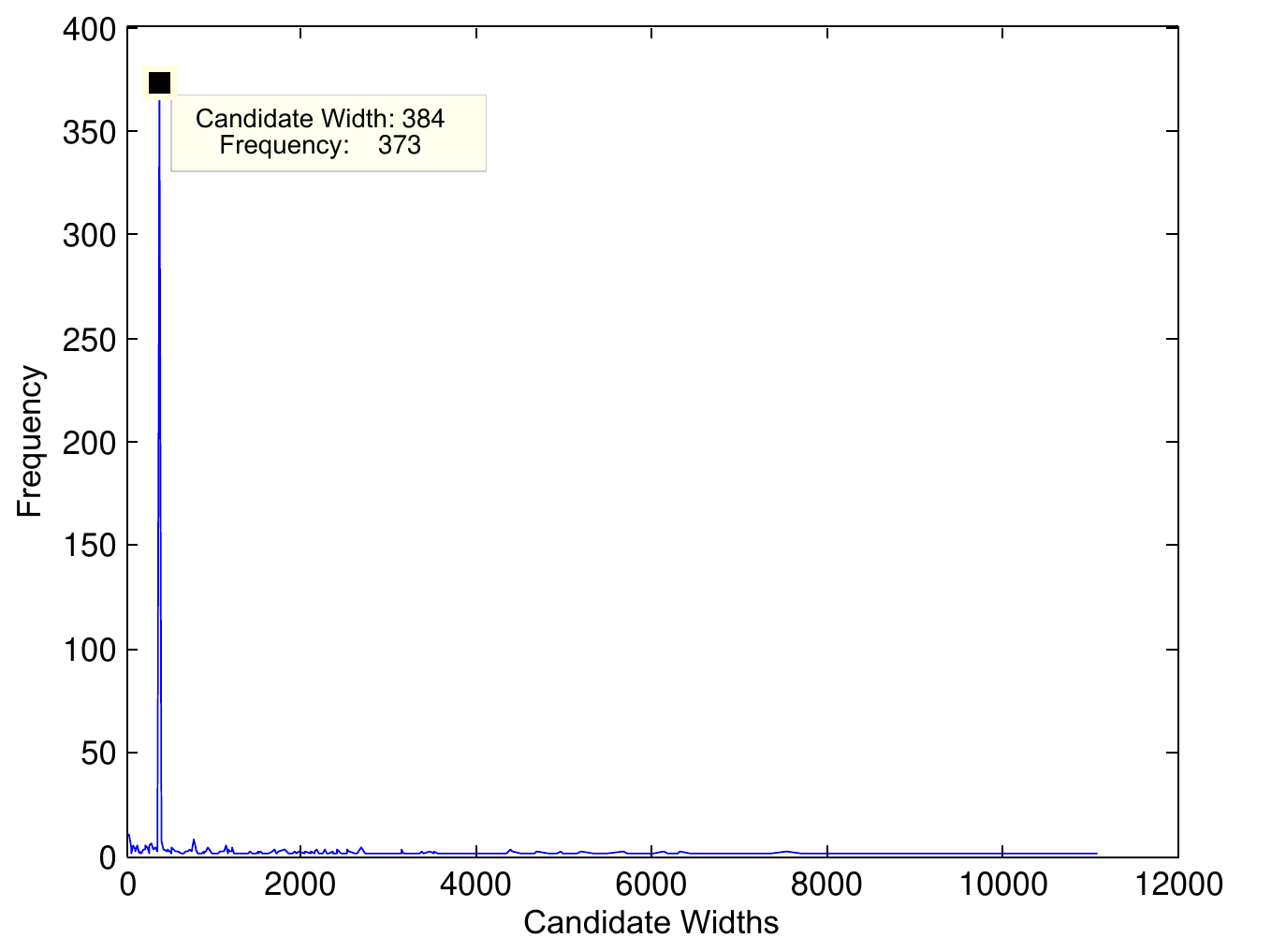}
\caption{\label{imageIntervalstatistic}  The result of counting the frequency of the candidate widths for the image file shown in Fig.\ref{fig:The images showed by wrong width}.}
\end{figure}

\begin{figure}[htbp]
\centering
\includegraphics[width=0.3\textwidth]{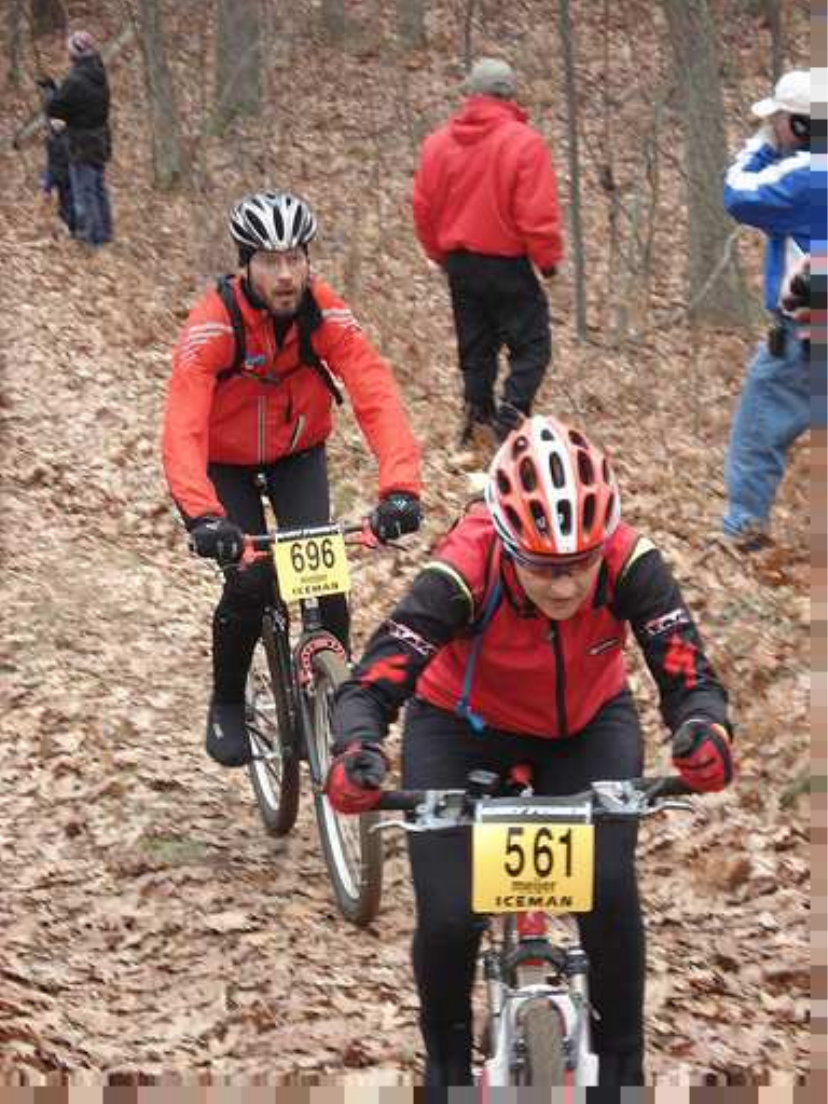}
\caption{\label{pascal_voc_10_2007_001630_estimation_width}  The decoded image data shown in Fig.\ref{fig:The images showed by wrong width} is displayed with the estimated width $384$ by proposed method.}
\end{figure}

%
We take the JPEG image file shown in Fig.\ref{fig:The images showed by wrong width} as an example to illustrate the proposed method.
768 MCUs in total are decoded from the JPEG file. The frequency distribution of the calculated candidate widths is shown in Fig.\ref{imageIntervalstatistic}. It can be seen that the highest frequency is 373 and the corresponding candidate width is 384, hence the image width is estimated as 384. The decoded image data displayed with the width 384 is shown in Fig.\ref{pascal_voc_10_2007_001630_estimation_width}. It is obvious that the image content can be fully understood and the block shifting phenomenon has been disappeared.

It should be noted that the estimated width is not strictly equal to the true image width 375 and slight image blurring appears at the right and bottom sides of the shown image. Both these are because MCU is the smallest group of data units encoded, which makes JPEG coding requires the test image satisfying the condition that the size of the coded image must be a multiple of the MCU size. If the condition is not satisfied, the coded image will be expanded. Hence the width of decoded images must be multiple of the MCU size. If the true image size is available,  image processing softwares will truncate the decoded image size into the true image size. Therefore, in the view of the image content, the estimated image width is equivalent to the true image width.

\section{EXPERIMENTS }
\label{sec:EXPERIMENTS}
The proposed approach is evaluated on  USC-SIPI \footnote{http://sipi.usc.edu/database/} image database and the PASCAL VOC 2010 Challenge dataset.

The USC-SIPI image database is rich in image content, which is divided into four volumes based on the basic character of the pictures. The four volumes are Textures, Aerials, Miscellaneous and Sequences.
The Textures volume consists of 64 standard textures and 91 rotated textures, all monochrome. The Aerials volume contains 38 images: 37 color, 1 monochrome. The Miscellaneous volume consist of 16 color and 28 monochrome images, in total 44 images.  The Sequences volume contains 69 monochrome images in 4 sequences. The image widths and the corresponding image numbers are summarized in Table \ref{TB- USC-SIPI image width and the corresponding image number}. Besides, since all images in the database are stored in TIFF format, we convert them to JPEG format by imwrite function in matlab. It is noted that the conversion does not have any impact on the evaluation.

\begin{table}[htbp]
\caption{The image widths in USC-SIPI database and the corresponding image numbers }\label{TB- USC-SIPI image width and the corresponding image number}
\vspace*{2pt}
\centerline{
\begin{tabular}{l|c|c}
\hline
& Image width & Image number \\
\hline
\multicolumn{ 1}{r|}{Textures} &        512 &        130 \\
\multicolumn{ 1}{r|}{} &       1024 &         25 \\
\hline
\multicolumn{ 1}{r|}{Aerial} &        512 &         12 \\
\multicolumn{ 1}{r|}{} &       1024 &         25 \\
\multicolumn{ 1}{r|}{} &       2250 &          1 \\
\hline
\multicolumn{ 1}{r|}{Miscellaneous} &        256 &         14 \\
\multicolumn{ 1}{r|}{} &        512 &         26 \\
\multicolumn{ 1}{r|}{} &       1024 &          4 \\
\hline
\multicolumn{ 1}{r|}{Sequences} &        256 &         49 \\
\multicolumn{ 1}{r|}{} &        512 &         10 \\
\hline
\end{tabular} }
\end{table}

\begin{figure}[htbp]
\centering
\includegraphics[width=0.5\textwidth]{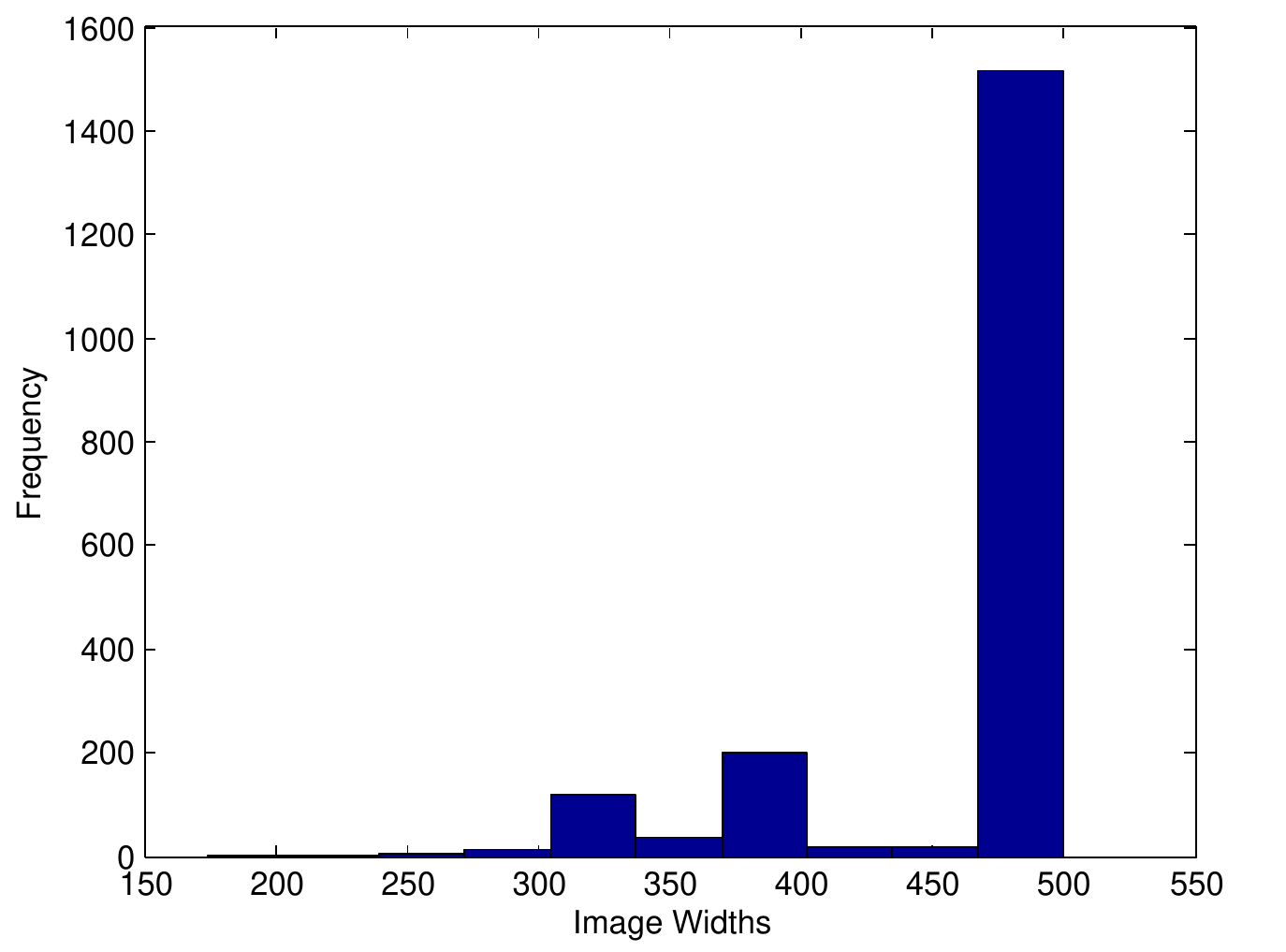}
\caption{\label{pascal_voc_10_wid_histogram} The widths histogram of PASCAL VOC 2010.}
\end{figure}

The PASCAL VOC 2010 \cite{everingham2010pascal} contains numerous real-world consumer images from Flickr\footnote{ http://www.flickr.com/}, and the public part of the dataset adding up to 1928 images are all JPEG format. The width histogram of this dataset is shown in Fig.\ref{pascal_voc_10_wid_histogram}. The minimum  and maximum image width in the dataset is $174$ and $500$ respectively, and 112 different widths in total.

In our experiments, we obtain the number of the components, sampling factors for all the components, Huffman tables and quantization tables from the header of JPEG files to decode the JPEG data into MCU sequences with correct order as shown in the top row in Fig.\ref{MCU}.

We measure performance with accuracy
 \begin{equation}
acc = \frac{M}{N}
\label{acc}
\end{equation}
where $N$ is the total of the test images and $M$ is the number of the images whose widths are correctly estimated. As stated in section  \ref{sec:PROPOSED METHOD}, although the estimated width $w^*$ is not strictly equal to the true image width ${w_{tr}}$ which is extracted from the JPEG file header,  it is still regarded as a correct estimation when $w^*$  is equal to ${\left\lceil {{w_{tr}}/K} \right\rceil *K}$, where$ \left\lceil  \cdot  \right\rceil $ is the ceiling function, and $K$ is MCU width as stated in Eq.(\ref{compute pixel interval}).

\begin{figure}[t]
\begin{minipage}[b]{0.21\linewidth}
  \centering
  \centerline{\includegraphics[width=1.2\textwidth]{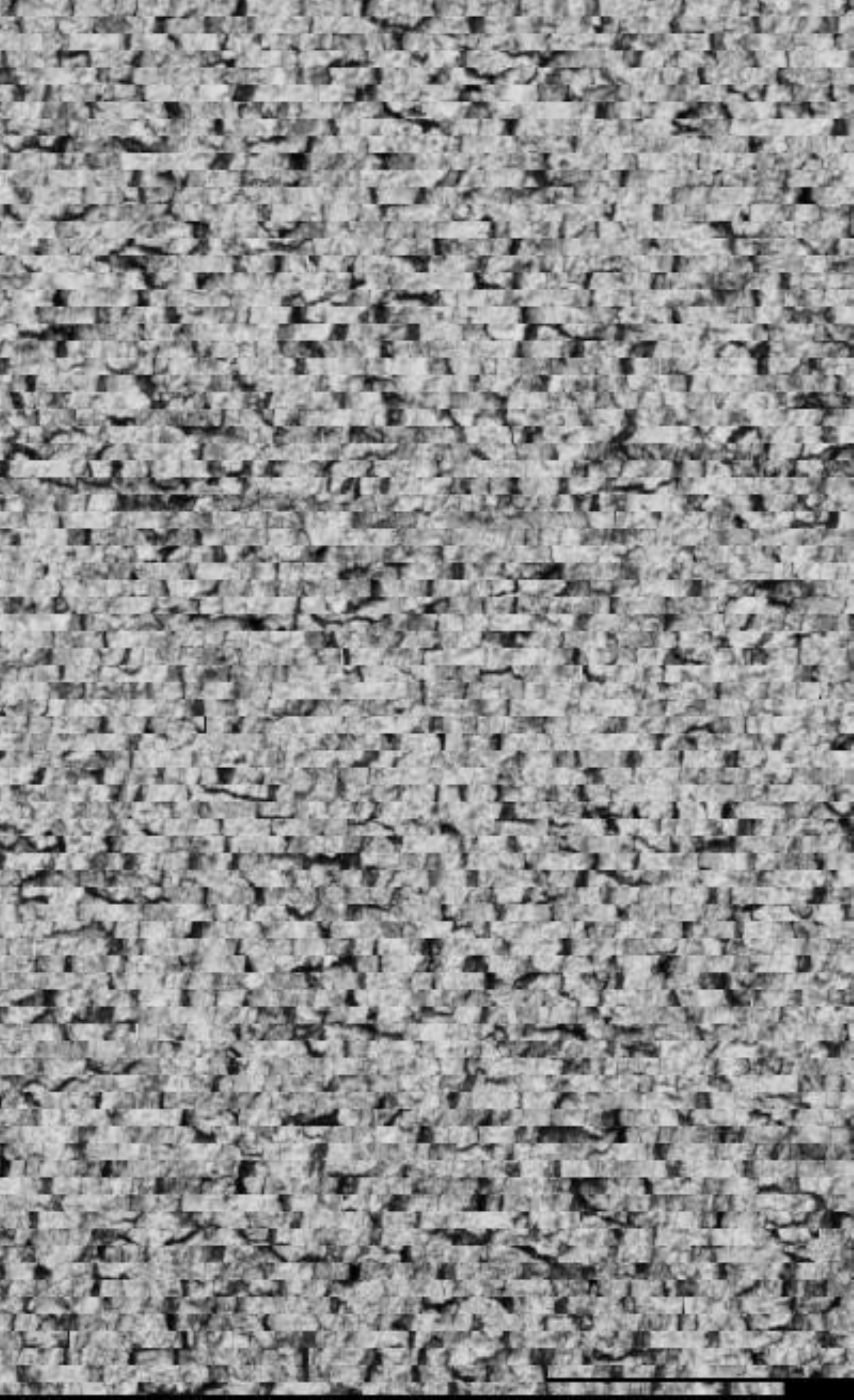}}
  \vspace{0.07cm}
\end{minipage}
\hfill
\begin{minipage}[b]{0.21\linewidth}
  \centering
  \centerline{\includegraphics[width=1.2\textwidth]{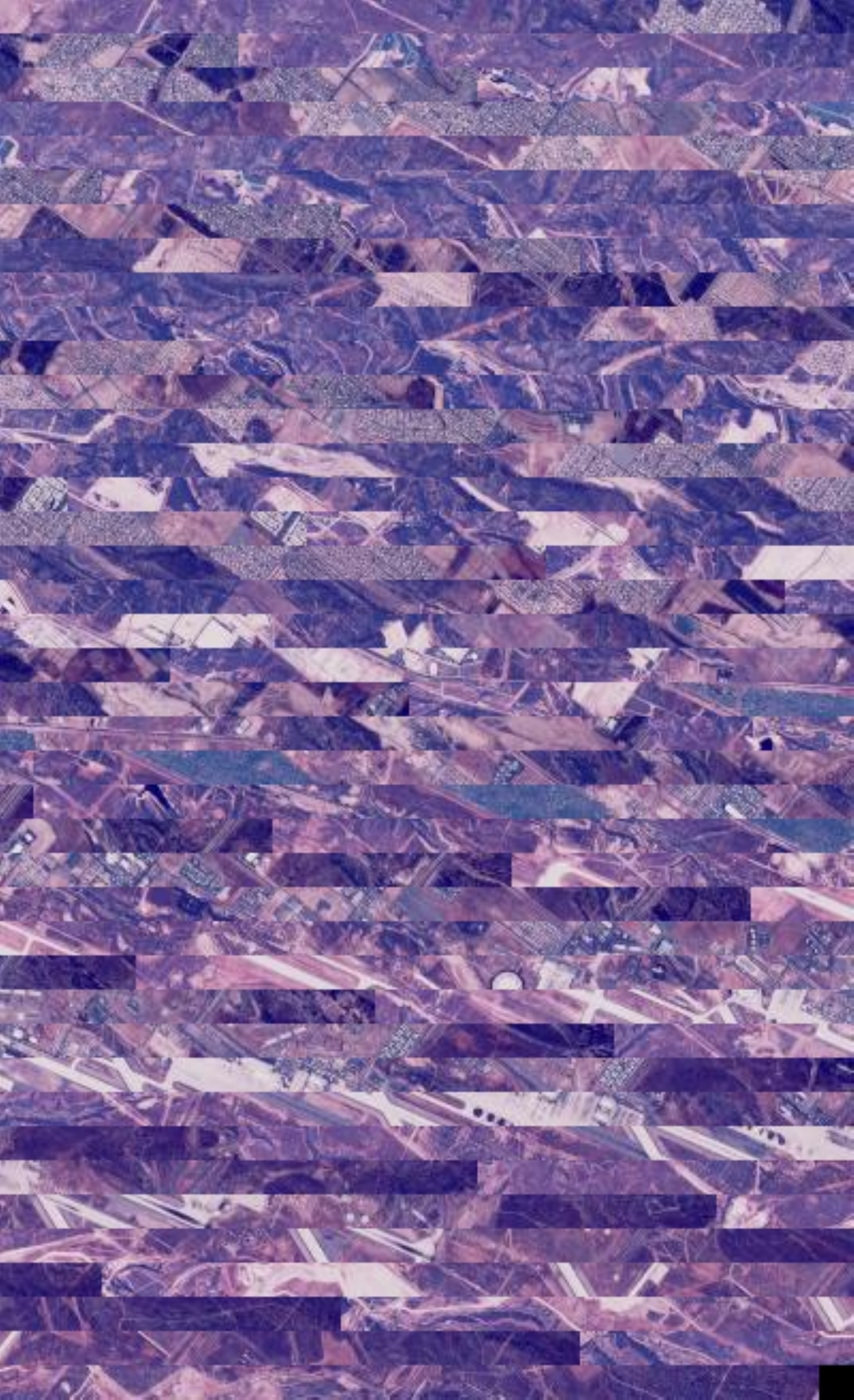}}
  \vspace{0.07cm}
\end{minipage}
\hfill
\begin{minipage}[b]{0.21\linewidth}
  \centering
  \centerline{\includegraphics[width=1.2\textwidth]{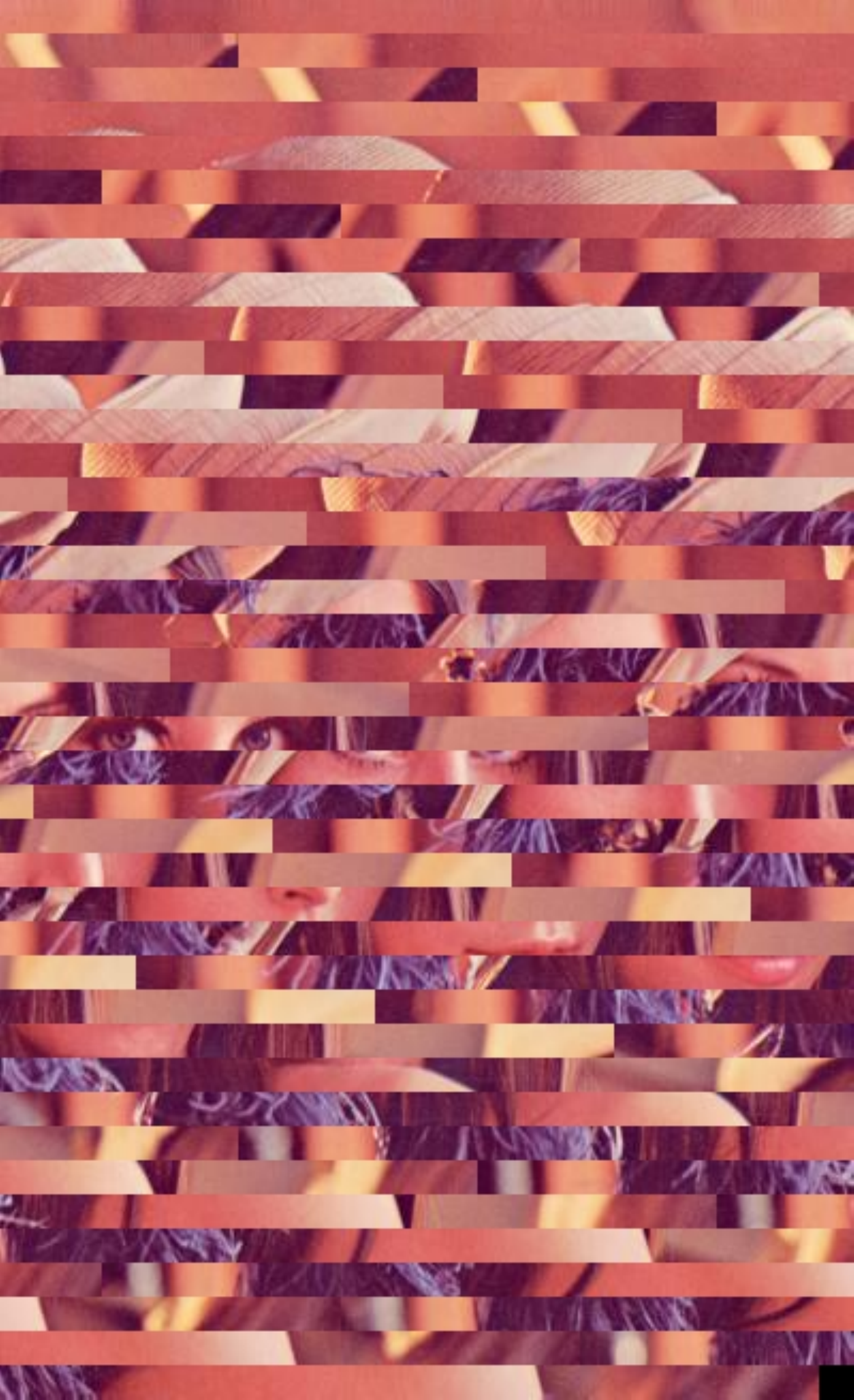}}
  \vspace{0.07cm}
\end{minipage}
\hfill
\begin{minipage}[b]{0.21\linewidth}
  \centering
  \centerline{\includegraphics[width=1.2\textwidth]{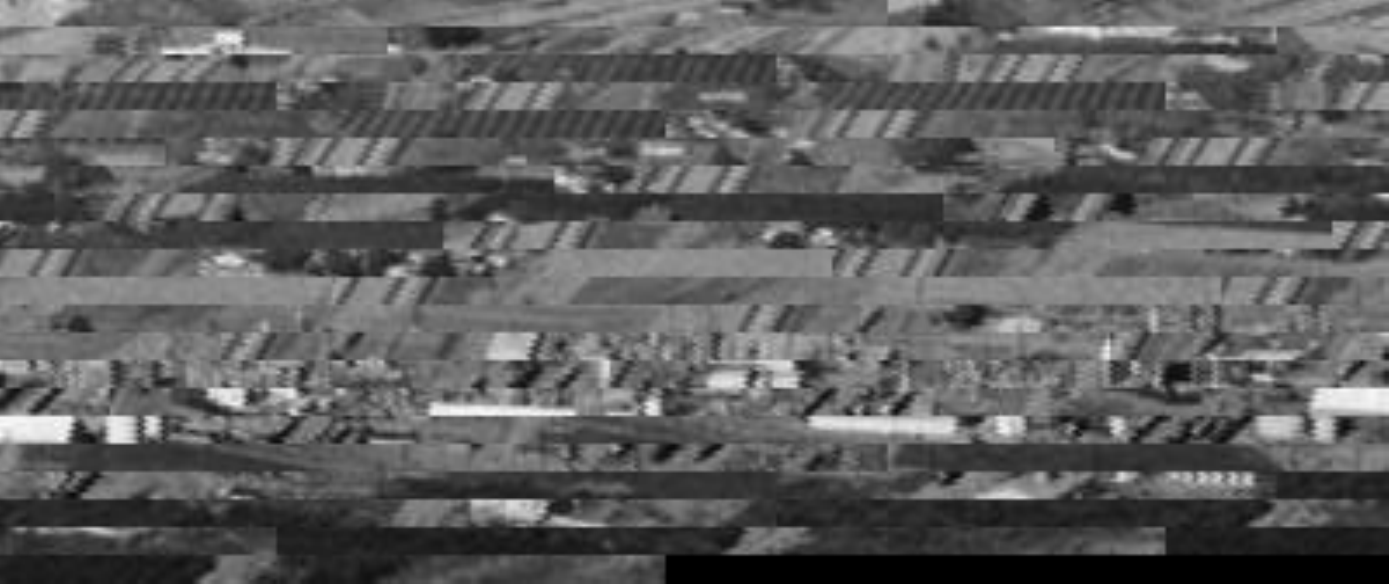}}
  \vspace{0.07cm}
\end{minipage}

\begin{minipage}[b]{0.21\linewidth}
  \centering
  \centerline{\includegraphics[width=1.2\textwidth]{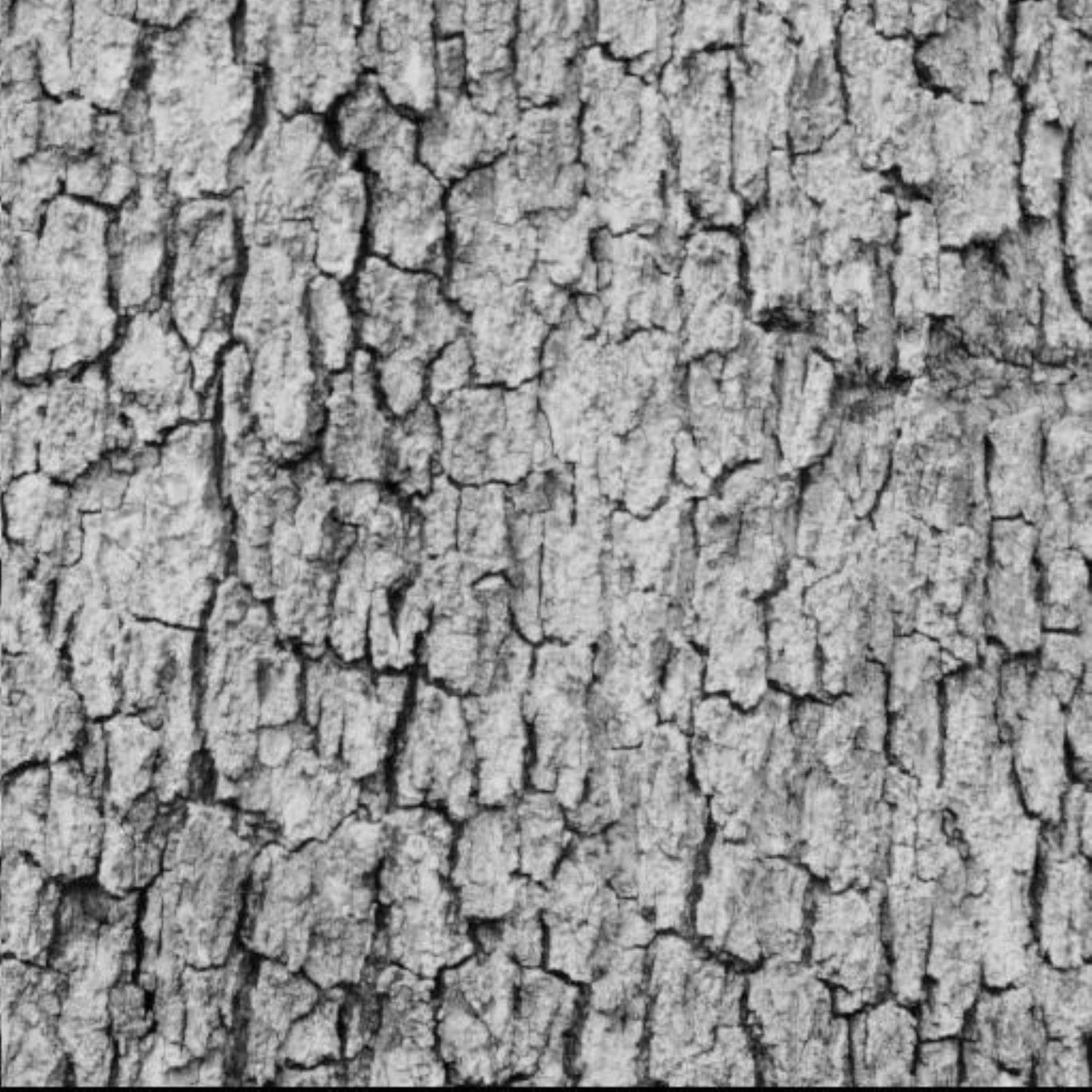}}
\end{minipage}
\hfill
\begin{minipage}[b]{0.21\linewidth}
  \centering
  \centerline{\includegraphics[width=1.2\textwidth]{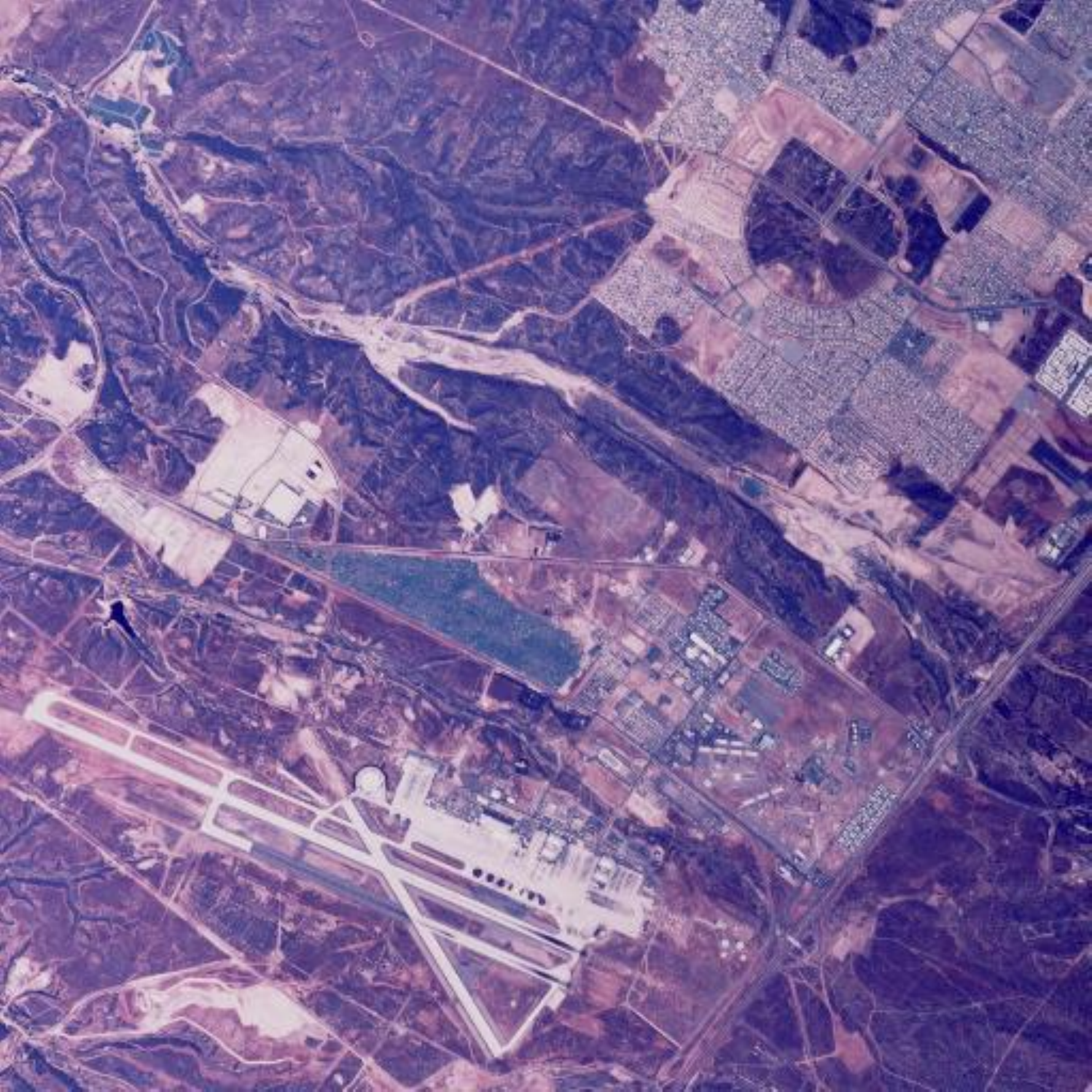}}
\end{minipage}
\hfill
\begin{minipage}[b]{0.21\linewidth}
  \centering
  \centerline{\includegraphics[width=1.2\textwidth]{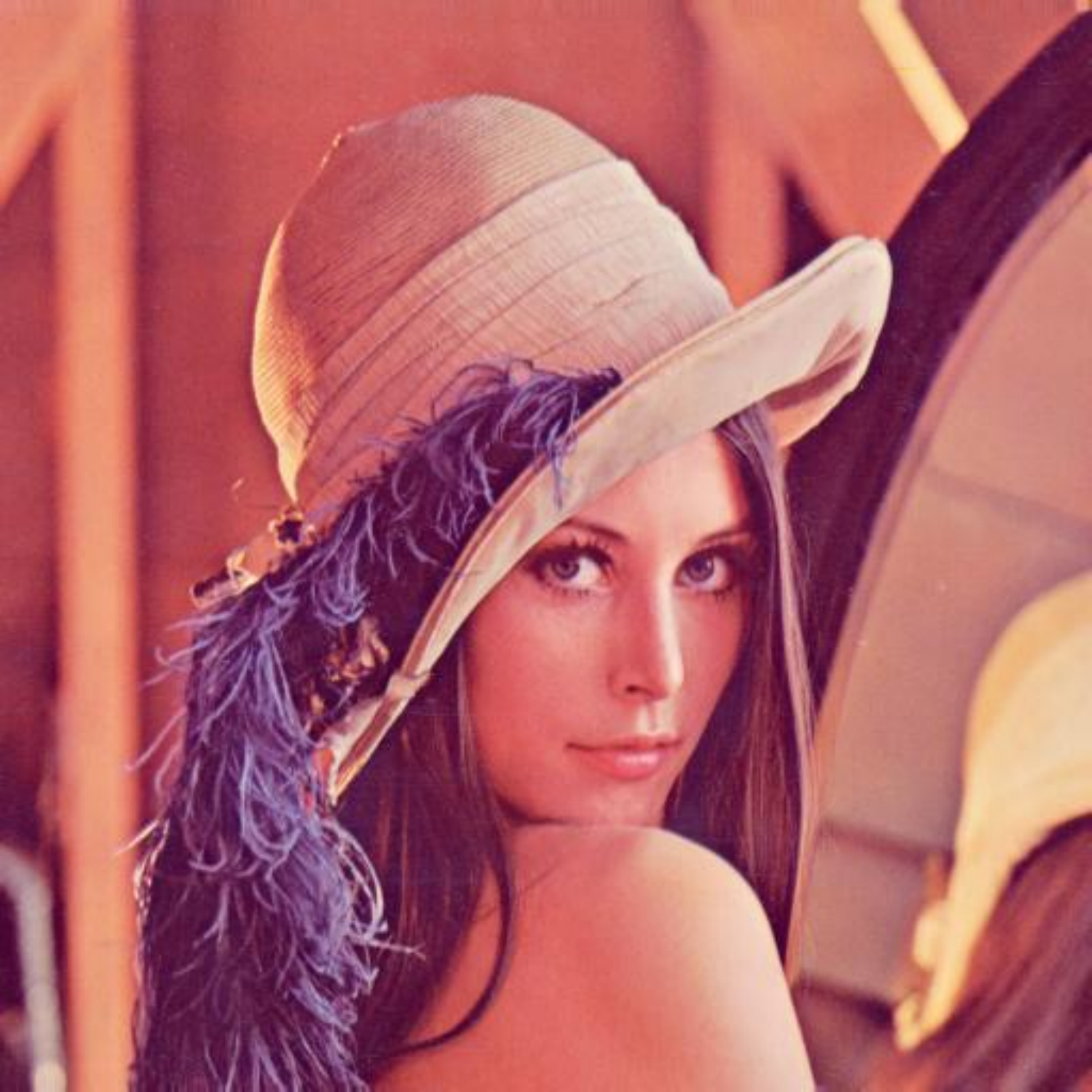}}
\end{minipage}
\hfill
\begin{minipage}[b]{0.21\linewidth}
  \centering
  \centerline{\includegraphics[width=1.2\textwidth]{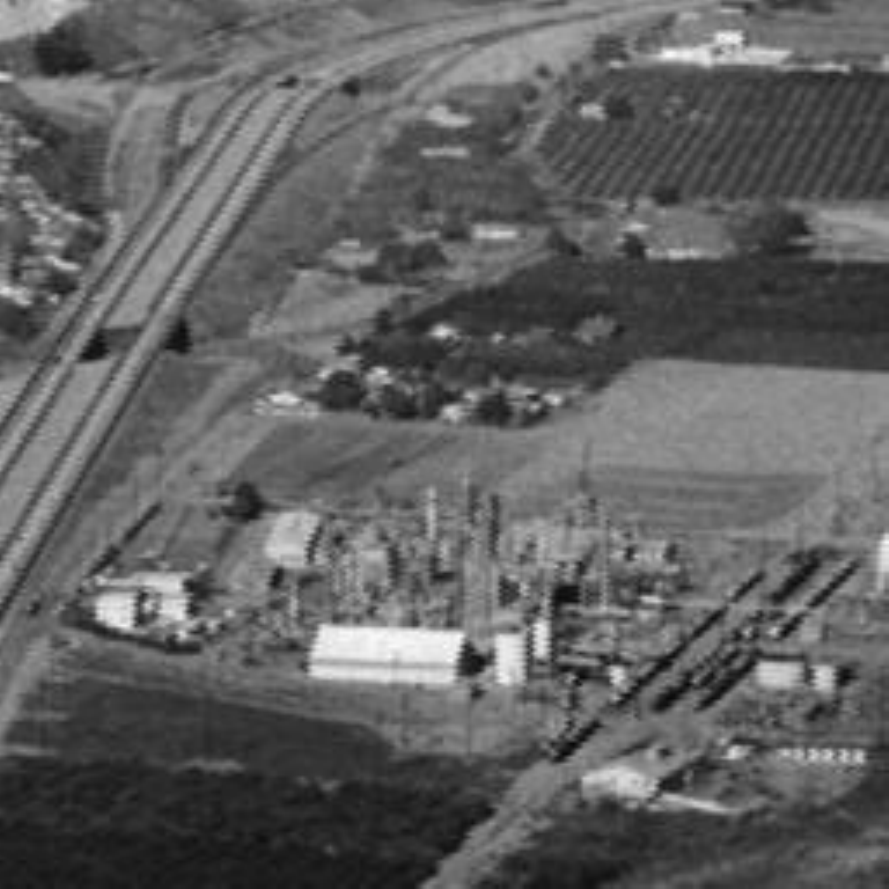}}
\end{minipage}

\caption{Example results from the USC-SIPI image database. Top row: four images displayed with the wrong width 400. Bottom: four images displayed with their estimated widths. These four images from left to right come from Textures, Aerials, Miscellaneous and Sequences respectively.
}
\label{fig:some instanse images SIPI}%
\end{figure}

\begin{figure}[t]
\begin{minipage}[b]{0.21\linewidth}
  \centering
  \centerline{\includegraphics[width=1.2\textwidth]{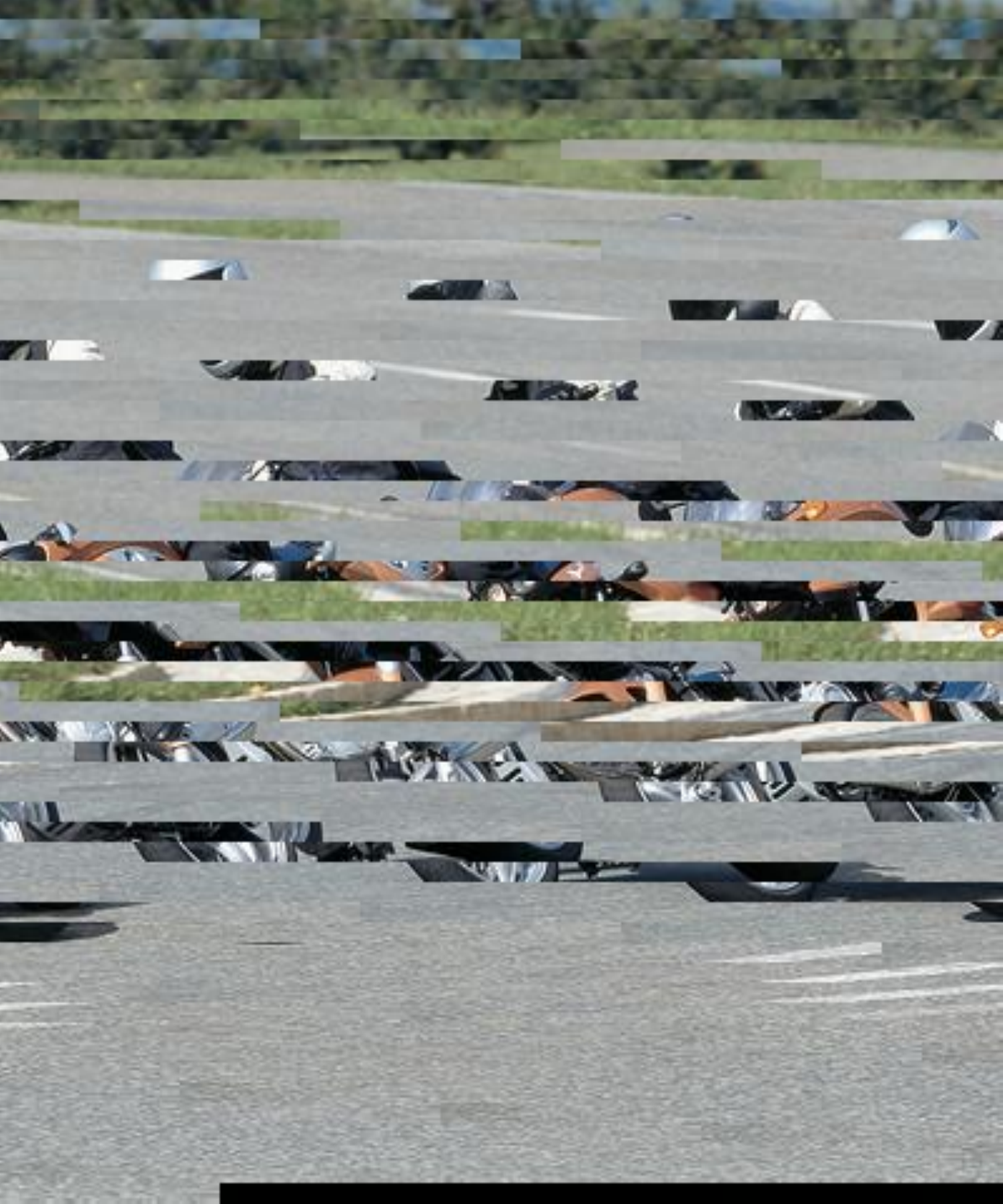}}
  \vspace{0.07cm}
\end{minipage}
\hfill
\begin{minipage}[b]{0.21\linewidth}
  \centering
  \centerline{\includegraphics[width=1.2\textwidth]{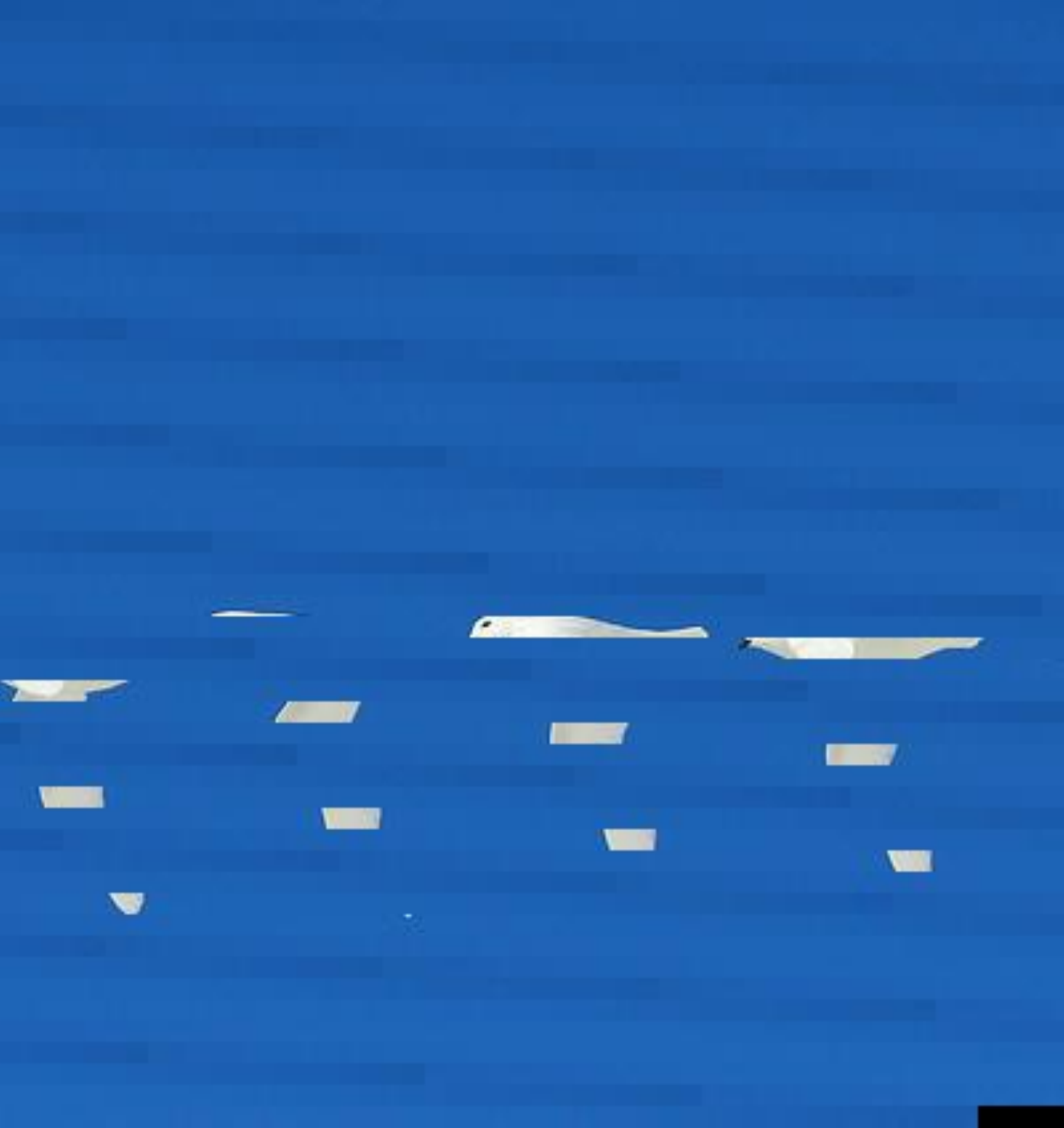}}
  \vspace{0.07cm}
\end{minipage}
\hfill
\begin{minipage}[b]{0.21\linewidth}
  \centering
  \centerline{\includegraphics[width=1.2\textwidth]{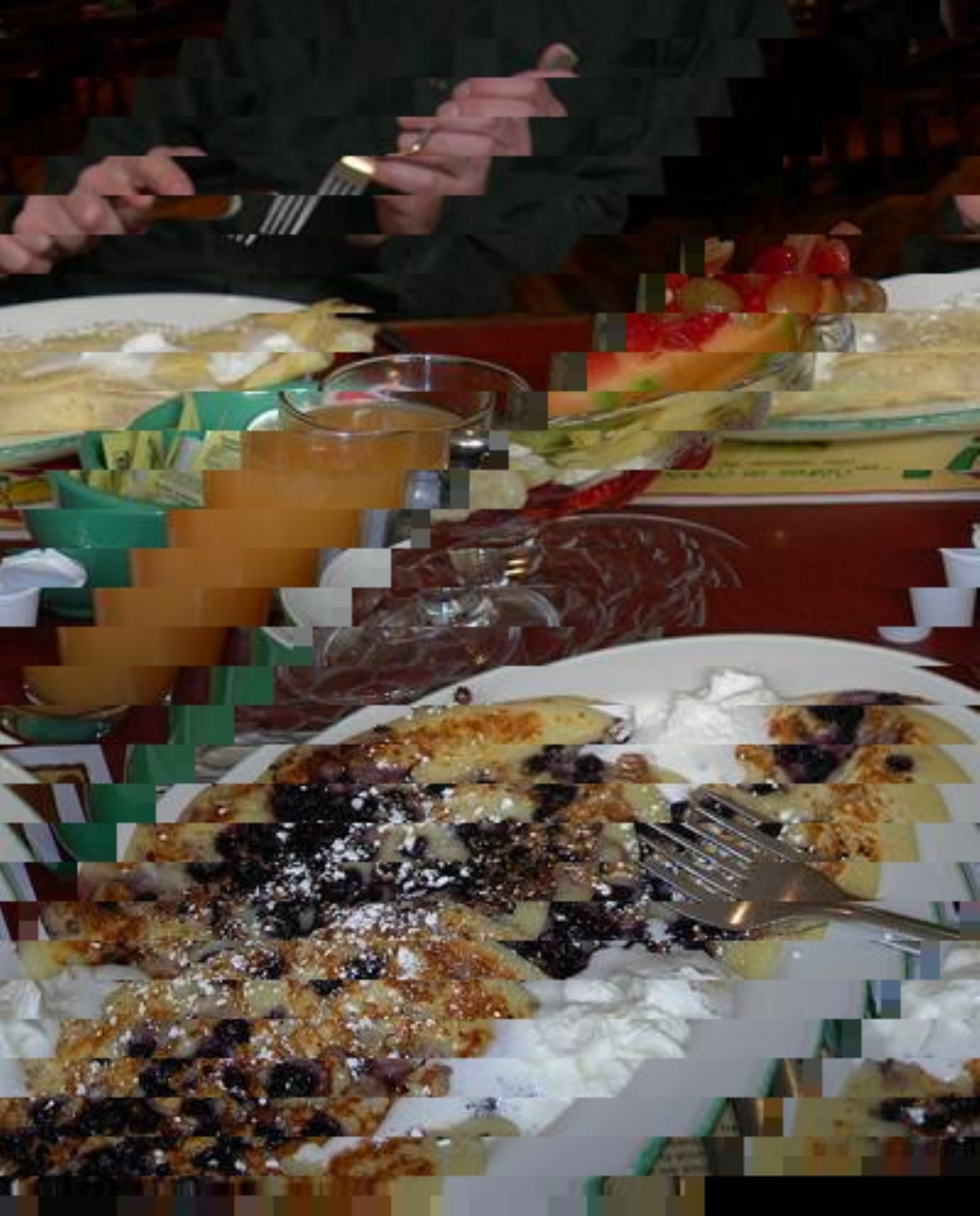}}
  \vspace{0.07cm}
\end{minipage}
\hfill
\begin{minipage}[b]{0.21\linewidth}
  \centering
  \centerline{\includegraphics[width=1.2\textwidth]{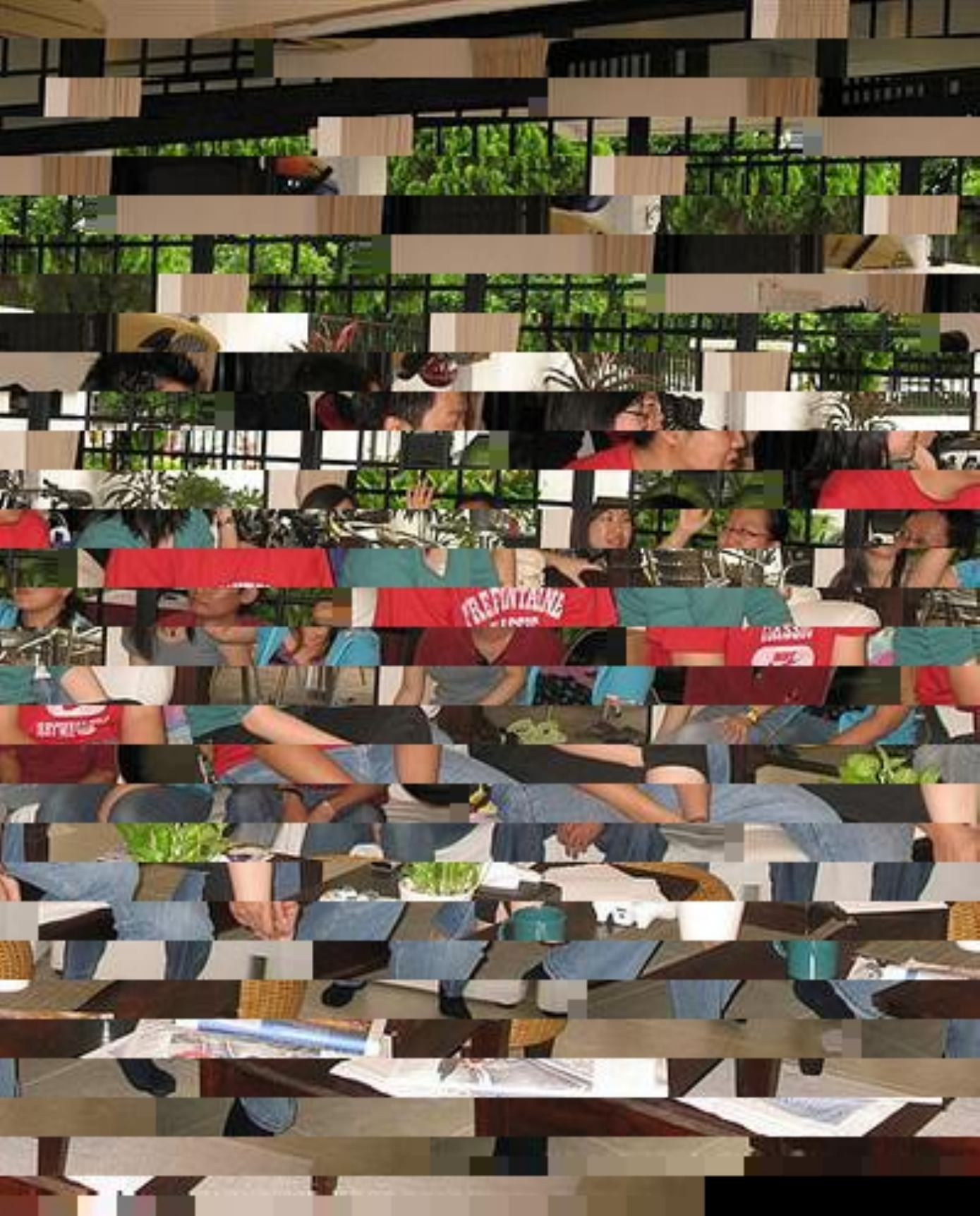}}
  \vspace{0.07cm}
\end{minipage}

\begin{minipage}[b]{0.21\linewidth}
  \centering
  \centerline{\includegraphics[width=1.2\textwidth]{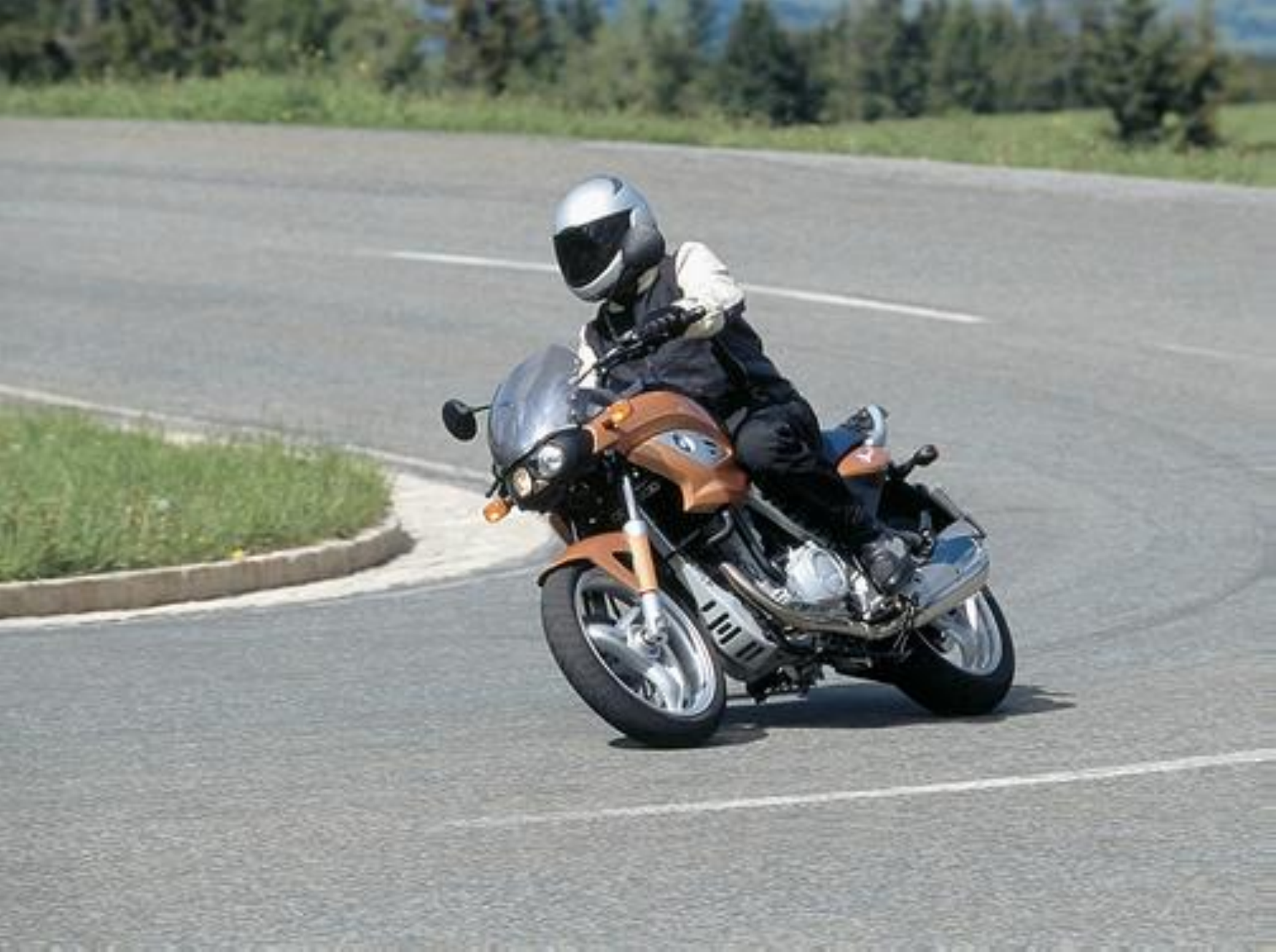}}
\end{minipage}
\hfill
\begin{minipage}[b]{0.21\linewidth}
  \centering
  \centerline{\includegraphics[width=1.2\textwidth]{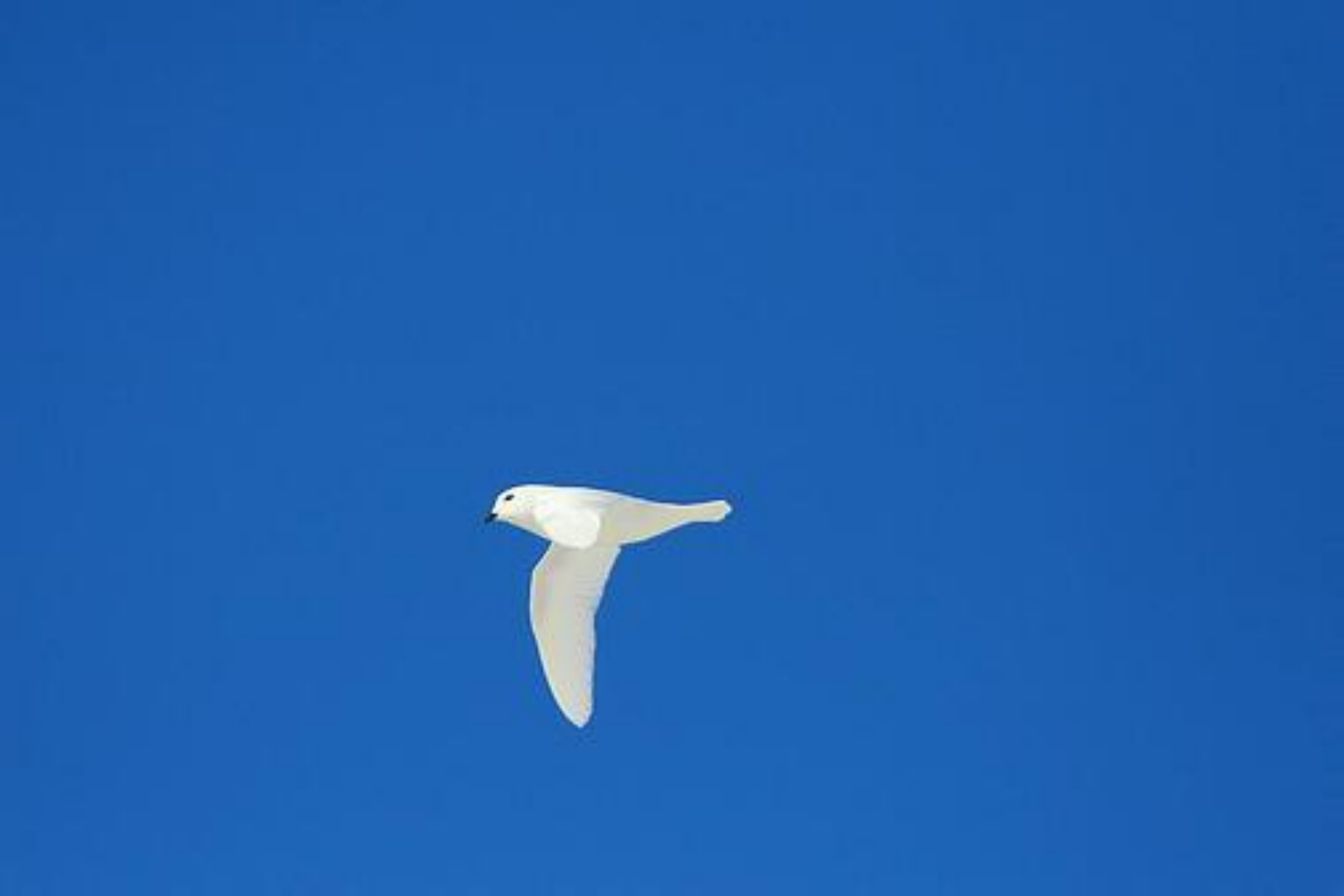}}
\end{minipage}
\hfill
\begin{minipage}[b]{0.21\linewidth}
  \centering
  \centerline{\includegraphics[width=1.2\textwidth]{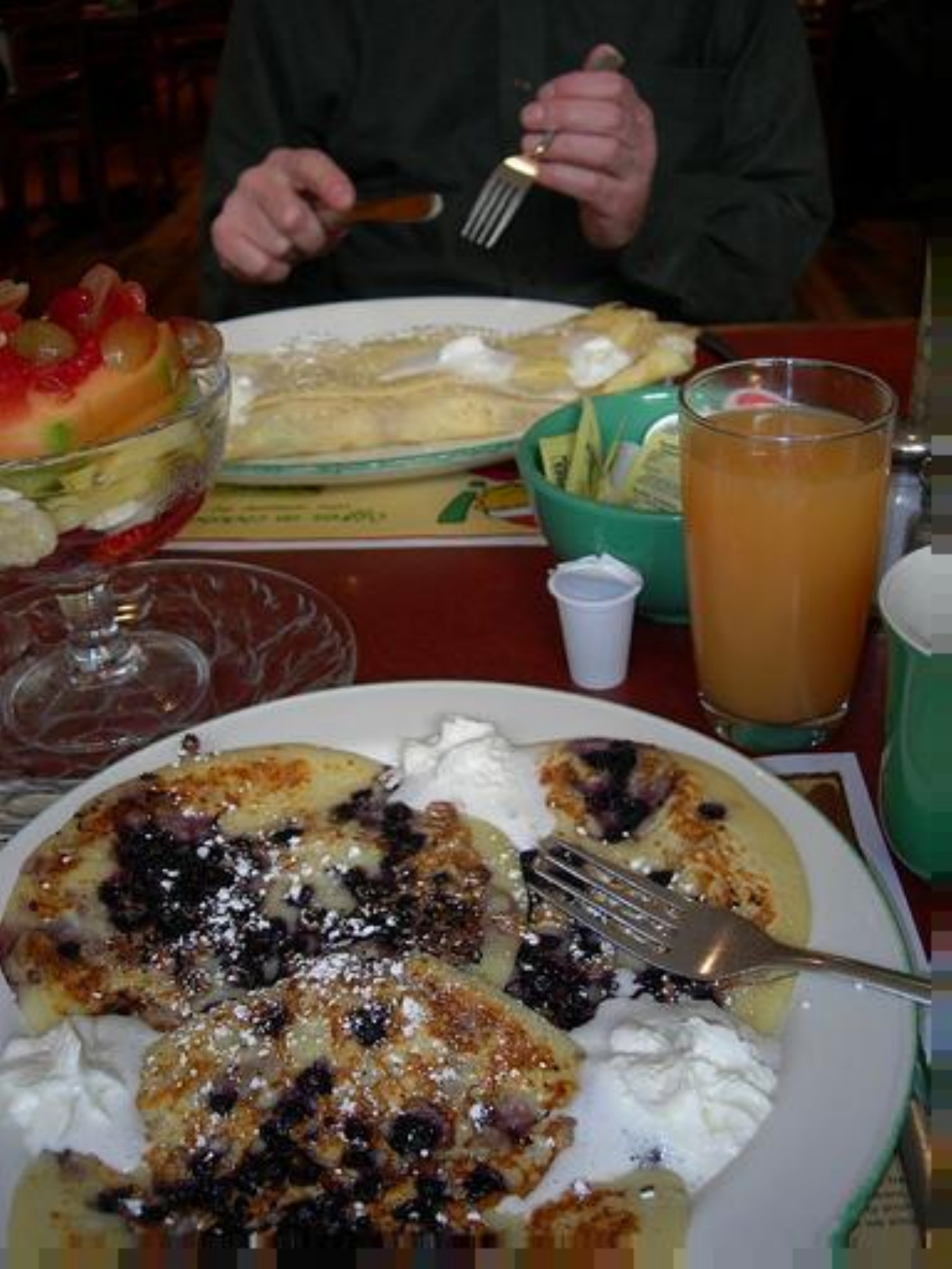}}
\end{minipage}
\hfill
\begin{minipage}[b]{0.21\linewidth}
  \centering
  \centerline{\includegraphics[width=1.2\textwidth]{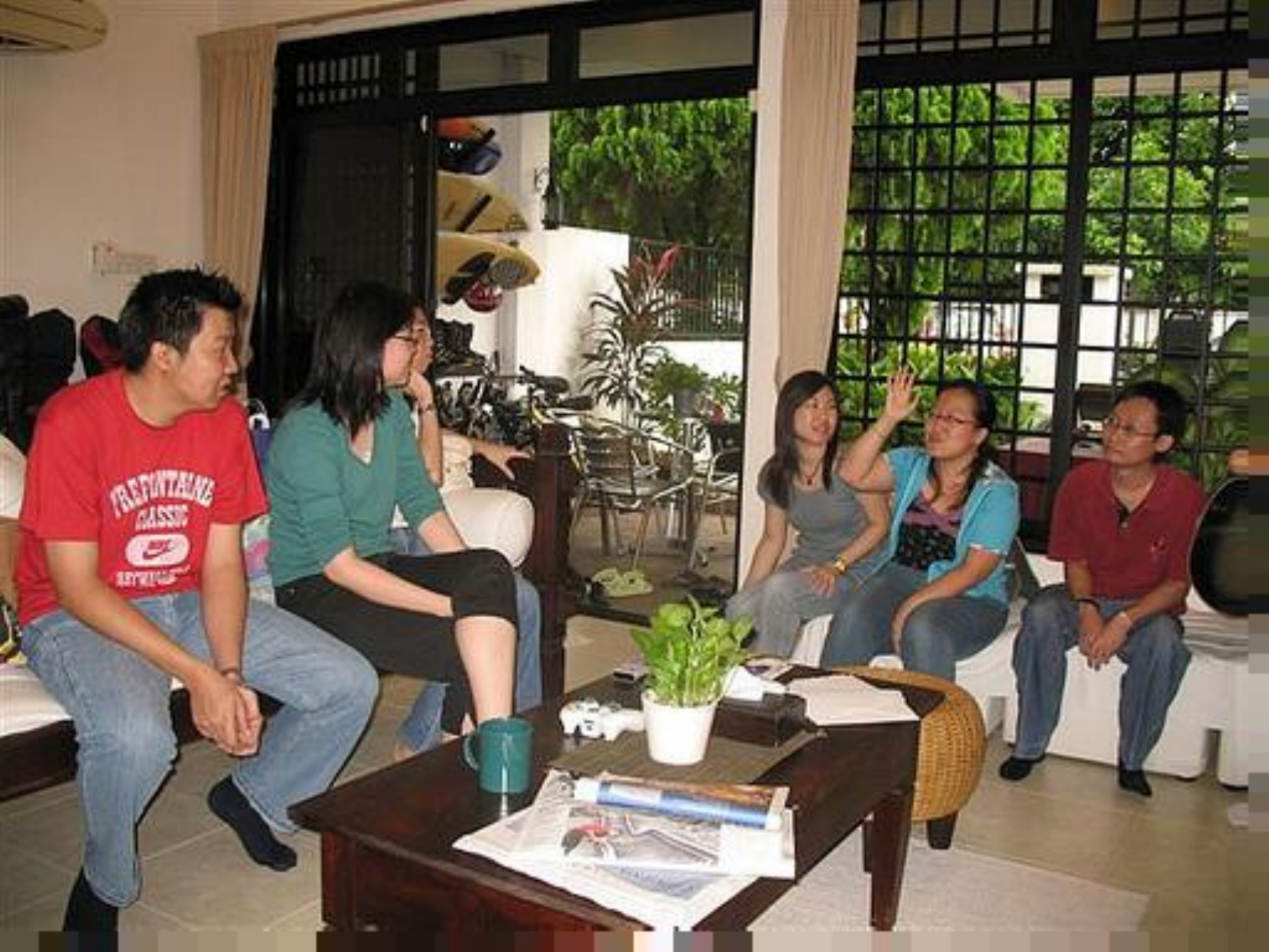}}
\end{minipage}

\caption{Example results from the PASCAL VOC 10 dataset. Top row: four images displayed with the wrong width 400. Bottom: four images displayed with their estimated widths.
}
\label{fig:some instanse images voc10}%
\end{figure}

The acc defined in Eq.(\ref{acc}) over  USC-SIPI image database and Pascal VOC10 is shown in Table \ref{TB-  experiment results on two imagedataset}, confirming the high performance of our method. Our method achieves  $100\%$ on Pascal VOC10 dataset, which means the widths of all the images from this dataset are correctly estimated. On USC-SIPI image database, 100\% are achieved for Sequences and Aerial volumes, while 99.35\% and 97.73\% are achieved for Textures and Miscellaneous volumes. Some example results from the two datasets are shown in Fig.\ref{fig:some instanse images SIPI} and Fig.\ref{fig:some instanse images voc10}. It is obvious that wrong widths have an serious impact on image understanding, and the proposed method can cope with this problem effectively.

\begin{table}[htbp]
\caption{The experiment results on USC-SIPI image database and Pascal VOC10}\label{TB-  experiment results on two imagedataset}
\vspace*{2pt}
\centerline{
\begin{tabular}{r|r|r|r}
\hline
           &          N &          M &        acc(\%) \\
\hline
Textures &        155 &        154 &    99.35 \\
Aerial &         38 &         38 &      100 \\
Miscellaneous &         44 &         43 &    97.73 \\
Sequences &         69 &         69 &      100 \\
Pascal VOC10 &       1928 &       1928 &      100 \\
\hline
\end{tabular}
 }
\end{table}

\begin{figure}[t]
\begin{minipage}[b]{0.45\linewidth}
  \centering
  \centerline{\includegraphics[width=1.1\textwidth]{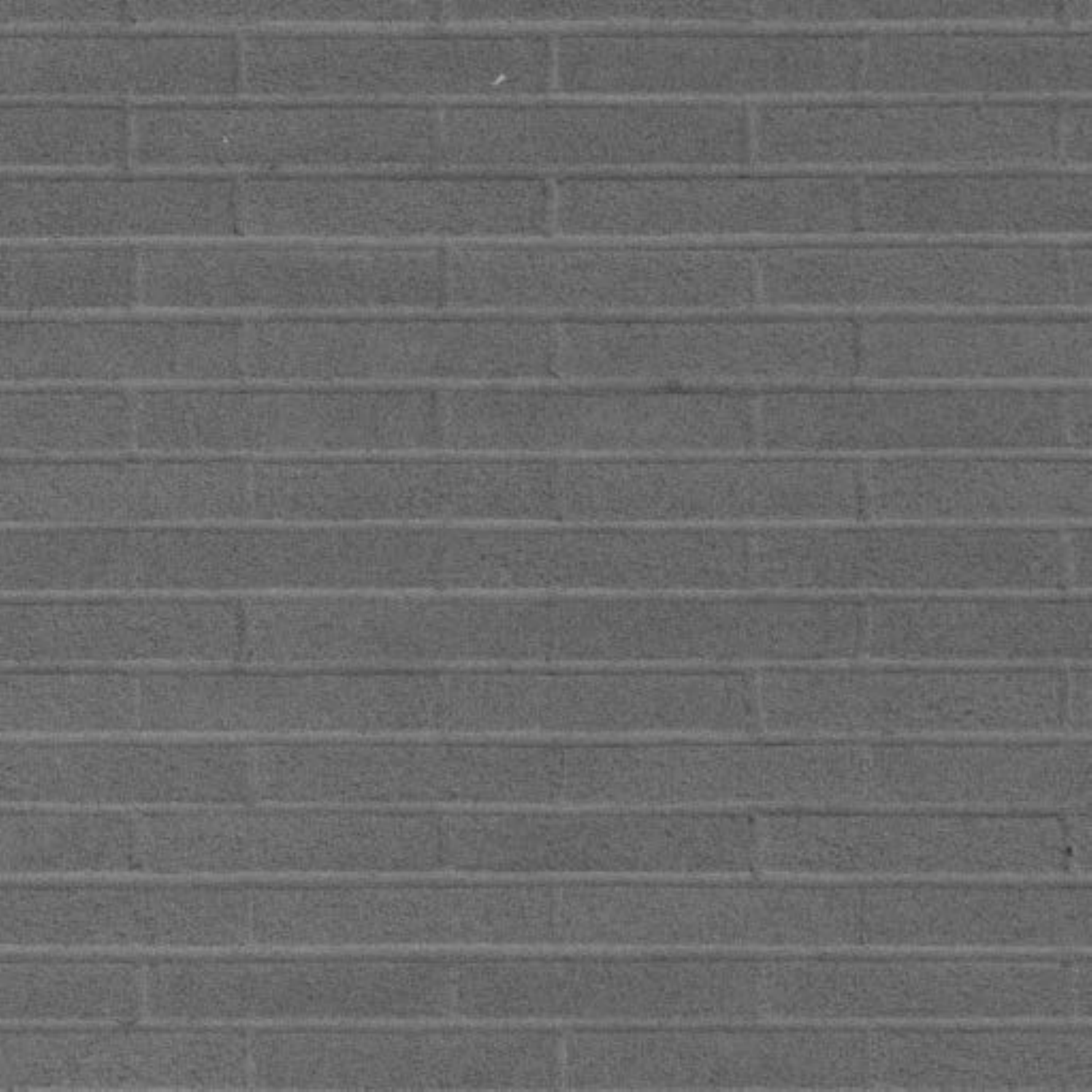}}
  \centerline{(a)}\medskip
\end{minipage}
\hfill
\begin{minipage}[b]{0.45\linewidth}
  \centering
  \centerline{\includegraphics[width=1.1\textwidth]{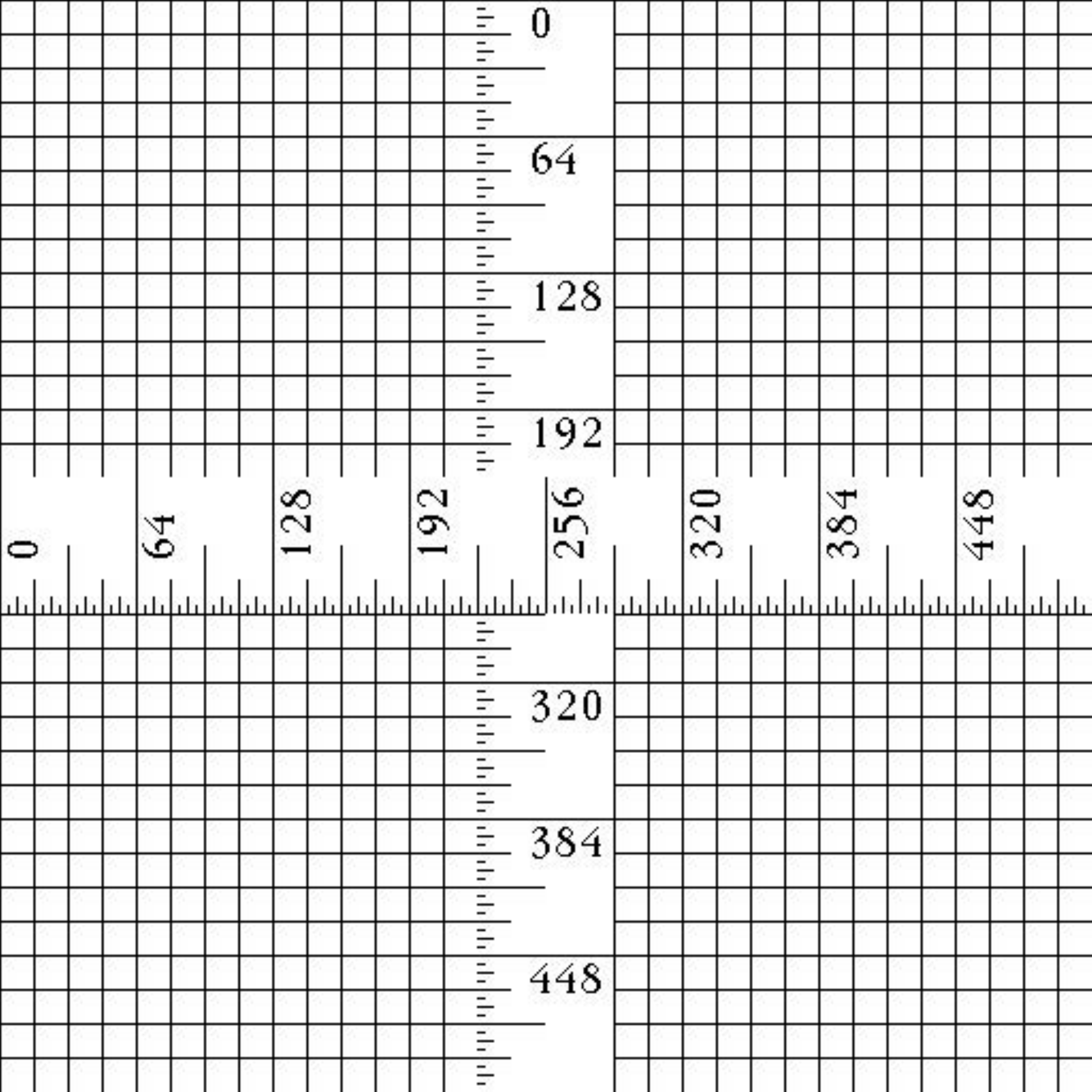}}
  \centerline{(b)}\medskip
\end{minipage}

\begin{minipage}[b]{0.45\linewidth}
  \centering
  \centerline{\includegraphics[width=1.1\textwidth]{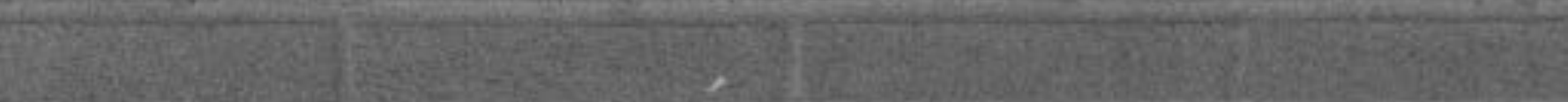}}
  \centerline{(c)}\medskip
\end{minipage}
\hfill
\begin{minipage}[b]{0.45\linewidth}
  \centering
  \centerline{\includegraphics[width=1.1\textwidth]{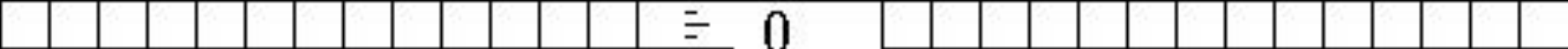}}
  \centerline{(d)}\medskip
\end{minipage}

\caption{The miscalculation images. (a): \emph{1.5.06.tiff} coming from Textures volume. (b): \emph{ruler.512.tiff} coming from Miscellaneous volume.(c): part of (a). (d): part of (b).
}
\label{fig:The miscalculation images}%
\end{figure}

The miscalculation image of Textures and Miscellaneous volumes contained in USC-SIPI image database is \emph{1.5.06.tiff} and \emph{ruler.512.tiff} respectively, which are shown in Fig.\ref{fig:The miscalculation images}. It can be observed that most pixels of the two images approximately have a certain periodicity. In other word, the images shown in (a) and (b) can be reconstructed approximately by repeating the figures (c) and (d) along the vertical direction respectively. The periodicity may be the reason that our method fails to work for the two images.

\section{CONCLUSION}
\label{sec:CONCLUSION}

We have presented a novel technique to estimate width for JPEG images when their widths is not available, whose basic idea is to find MCU pairs adjacent in the vertical direction according to the local similarity of images. The local  similarity is measured by the average Euclidean
distance between the pixels from the bottom row of the top MCU and the top row of the bottom MCU. Experimental results show that the proposed method achieve very high accuracy on PASCAL VOC 2010 challenge and USC-SIPI image database, and the widths of almost all these test images are estimated correctly.

%
%
%
\bibliographystyle{IEEEbib}
\bibliography{refs}

\end{document}